# Directional Control of Droplet Motion via Geometric Gating

Manoj K. Chaudhury[1], Vanessa Elias and Michelle Fruge

Chemical and Biomolecular Engineering, Lehigh University
Bethlehem, PA 18015 (USA)

**Abstract.** We explore new methods to induce motion of water droplets on elastomeric substrates, such as polydimethylsiloxane (PDMS). Our approach builds on the well-known phenomenon that water droplets move easily across these substrates when swelled with silicone oil. While designing shallow open channels and traps on a PDMS film offers a simple way to control a droplet's trajectory by tilting the substrate in two orthogonal directions, this method is poorly suited for manipulating multiple droplets simultaneously. To address this limitation, we introduce two new methods. The first method uses a plano-concave PDMS platform with a radius of curvature much larger than the droplet size and its translation distance. When the system configuration changes, this curved substrate consistently guides the droplet along a geodesic path. The second method incorporates diode-like gates on the PDMS. We create these gates on either flat or plano-concave PDMS films by embedding thin, rigid plates into the elastomer before crosslinking, then swelling in silicone oil. This process generates sigmoidal steps that restrict droplet motion to a single direction. Finally, we demonstrate that combining these two strategies enables various forms of directed motion for both single and multiple droplets.

[1]mkc4@lehigh.edu

## Introduction

While droplet microfluidics has traditionally focused on the controlled routing of microliter droplets through closed channels[1,2], the field is expanding. Over the past few decades, significant research has aimed to achieve similar control in open-surface microfluidics by harnessing forces such as modulated capillarity, inertia, magnetism, electricity, and electrocapillarity.[3–11] Although these methods often require complex fabrication, they have achieved varying degrees of success. Across these approaches, contact angle hysteresis—whether pre-existing or developed during fluidic operation—is widely recognized as a major impediment to droplet motion and a key hurdle to building reliable open-surface microfluidic devices. This topic has been extensively discussed in several reviews [3-11] and remains an active area of research, as evidenced by a growing number of recent publications.

While generating droplet motion on a surface is the fundamental requirement in open-surface microfluidics, achieving directed transport and integrating various unit operations on a single device remain long-term objectives. Recently, the field of microfluidics, including open-surface systems, has been enriched by automation engineers who have made extraordinary progress deploying AI-based methods[12-14], such as deep learning, reinforcement learning, and digital twins. Alongside these developments, paradigm-shifting research has emerged around the concept of "autonomous" droplets—liquids capable of sensing their environment and making autonomous adjustments to navigate on their own[15,16].

We previously explored how droplet motion can be generated on a substrate subjected to asymmetric vibration[17], a mechanism in which contact angle hysteresis plays a central role in symmetry breaking. Leveraging this concept, Daniel[17] developed several prototype devices that guided discrete droplets across an open surface to various unit operations. This setup used a computer program to precisely automate the direction, speed, and distance traveled by specific drops. Beyond this approach, several other techniques have demonstrated—or show strong potential for—the controlled guidance of liquid droplets on open surfaces. For instance, Mertaniemi et al.[18] used superhydrophobic tracks with negligible contact angle hysteresis to drive droplet motion via slight substrate tilting. Programmable metamaterial-induced[19-21] localized vibration has also emerged as a viable method for creating directed droplet movement, while charge[7]- and elastocapillary[22]-induced droplet gating also hold significant promise. We will return to the latter study, which may bear on what we present here.

When using droplet-based fluidic devices for temperature-sensitive chemical reactions, we must keep in mind that droplets evaporate at elevated temperatures—particularly near the boiling point. This poses a critical challenge. To mitigate this, fluid loss can be countered[17] by intermittently supplying additional water to the droplet. Alternatively, oil-encapsulated droplets can significantly reduce evaporation. About a decade ago, we developed a method[23] to induce microdroplet motion on an oil-covered elastomeric film impregnated with a square matrix of ferromagnetic particles. In this configuration, activating a specific particle with an electromagnet induces a localized deformation in the film. A nearby water droplet experiences an unbalanced Laplace force, driving it toward the center of the deformation. By sequentially addressing the particle array, the droplet can be precisely translated across various lattice points over long distances. Because the thin underlying oil layer separates the droplet from the substrate, contact angle hysteresis is mitigated, enabling uninhibited motion in any direction. Furthermore, the oil wetting the base of the droplet suppresses evaporation, allowing the system to be safely heated to study temperature-induced phase changes at approximately 60 °C.

Because a water droplet can move on a silicone oil-swollen polydimethylsiloxane (PDMS) elastomer even at a $1^0$ tilt and can be steered by controlling the tilt in two orthogonal directions, we initially wondered whether a droplet could be manipulated across a flat substrate without additional constraints. While this goal is largely achievable for a single droplet, it often deviates from its intended path. This deviation occurs because the oil puddle around the drop interacts with the background oil film, whose thickness becomes non-uniform after tilt-induced flow and after previous drops pass. Although these effects are subtle, they generate a non-uniform, Laplace-pressure-gradient-driven drift in an uncontrolled direction. While such a drift may be ignored for crude droplet manipulation, it becomes a significant concern for high-precision control. Furthermore, although meticulous orthogonal tilting can manage a single droplet's trajectory, independently controlling multiple droplets under these conditions remains unfeasible.

While precisely manipulating a droplet's trajectory on a flat, featureless substrate is difficult, various geometric structures can be integrated into the PDMS surface, building on a theme introduced by Mertaniemi et al.[18]. For instance, Elias and Fruge[24] used a silicone oil-coated PDMS substrate with shallow open channels and traps, providing excellent control over droplet trajectories. By using a hand-built, electromagnetically controlled stage to tilt the substrate in two orthogonal directions, they guided and arrested droplet motion along these surface features (*vide infra*). A portion of a droplet could also be transferred between traps by splitting it with a hydrophobically modified glass needle coupled to a peristaltic pump. Controlling the flow rates enabled either symmetrical division via suction or the generation of sequential daughter droplets via controlled pinch-off instability. This dual-mode splitting enables precise reagent delivery and the unique capability to recover and recycle droplets for further analysis (such as UV spectroscopy)[24B].

Raufaste et al.[25] achieved controlled droplet motion on an oil-coated PDMS substrate by encapsulating a small ferromagnetic particle within the droplet and driving its motion with an external magnetic field. We anticipate future extension of this magnetic actuation method to manipulate multiple droplets simultaneously on a single PDMS platform.

In this work, we report two additional complementary methods for inducing controlled water droplet motion. The first method uses a plano-concave PDMS platform that provides a single gravitational potential minimum on its curved surface at any inclination. Consequently, a droplet initially placed at the center moves along a curved geodesic to a new stable equilibrium position when the system's configuration changes. Because the droplet traverses equilibrium positions along a geodesic bounded by longitudinal lines of higher gravitational potential—effectively creating artificial channels—no droplet runaway occurs during the transition between potential minima. In fact, with slow steering, the droplet remains in a nearly undifferentiated equilibrium state. This approach enables precise droplet translocation to various positions on the curved PDMS surface without external guidance.

External guidance can, however, be used to manipulate multiple droplets on a single platform. This requires developing a transport arrest mechanism—analogous to an electronic diode or a fluidic check valve—that traps specific droplets at designated surface locations while allowing others to move freely. Such a mechanism enables coordinated steering of multiple droplets to accomplish specific tasks. This paper summarizes our studies, focusing primarily on: (a) droplet motion on a slightly curved PDMS substrate, (b) the mechanism of droplet trapping, and (c) the manipulation of multiple droplets on a surface that combines these two properties.

**Experimental Section**

Plano-concave and flat PDMS films bonded to glass slides served as the foundational platforms for the studies presented here. We swelled these PDMS films in silicone oils of three viscosities to obtain hysteresis-free surfaces suitable for manipulating water droplets. We induced droplet motion by slightly tilting the substrate in two orthogonal directions using either an electromagnetically controlled platform (Figure 1A) or a motorized bi-axial tilting platform (Figure 1B). Foton probes[26] (Figure 1B) were used alongside standard optical telemicroscopes to track droplet motion and examine droplet profiles.

To heat the PDMS substrate—such as during the redox reaction between glucose and Benedict's reagent—we attached a thin-film heater to the back of the glass slide and powered it with an electrical current. We achieved diode-like droplet motion by creating sigmoidal gates in the PDMS through inhomogeneous film swelling.

The Results and Discussion section is structured so that readers can follow the primary findings without prior knowledge of the experimental methods. The only exception is the biaxial tilt apparatus, which was central to all studies reported below. The experimental procedures are described in detail in the Supplementary Materials.

**Apparatus For Bi-Axial Tilts**

We controlled the substrate's biaxial tilt in two ways. Initially, we used a hand-built, electromagnetically controlled tilt mechanism. We affixed four permanent magnets to the underside of a rectangular glass plate that supported the substrate. We positioned this assembly above four electromagnets, separated by a gap. To maintain this gap, we placed a second glass plate over the electromagnets, using spacers to isolate the upper and lower plates while ensuring the permanent magnets faced the electromagnets directly. By leveraging both magnetic repulsion and attraction between the two sets of magnets, we successfully controlled the tilt of the upper platform along two orthogonal axes. In later stages of our research, however, we replaced this setup with a motorized biaxial goniometer, which provided automated clockwise and counterclockwise tilting of the substrate.

The droplet-dispensing needle and the Foton probe were independently translated along the X, Y, and Z axes using triaxial manipulators. For the Foton probe, a hydraulic manipulator provided additional, finer control over its biaxial tilt and triaxial translation. Both the triaxial and hydraulic manipulators were operated via joysticks. Finally, two telescopic microscopes with video recording capabilities enabled real-time visualization of the droplets, needle, and Foton probe from two orthogonal directions.

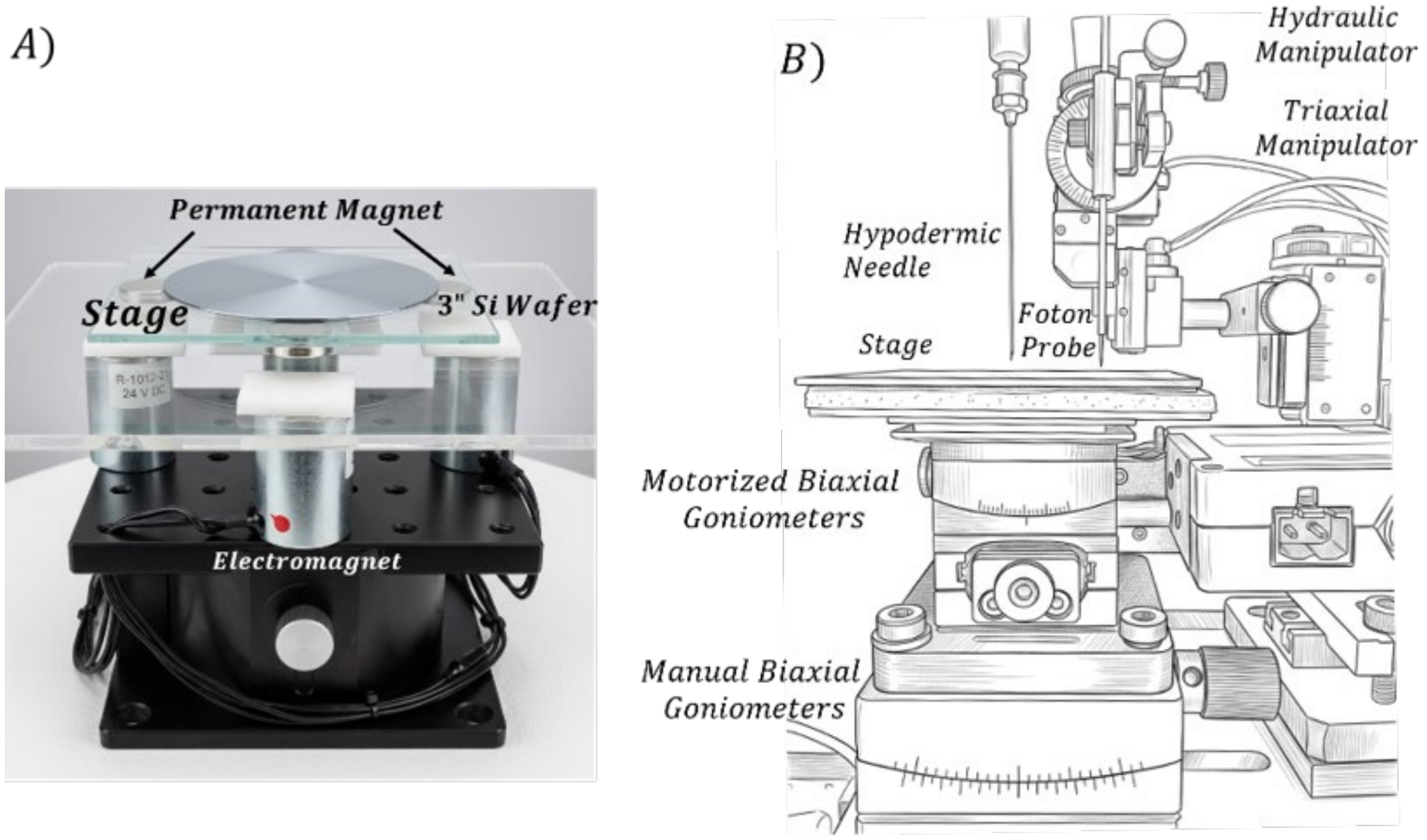

**Figure 1.** Schematics of the apparatus used to generate droplet motion on a substrate. A) Electromagnetically Controlled Tilt Apparatus. B) Tilt generated by biaxial goniometers. The figure identifies different components of the apparatus. See also Supporting Supplementary Materials: Figure SM1.

## Results and Discussion

### Guiding Droplet Motion Via Surface Structures

In the early stages of our research, we used a silicone oil-coated PDMS substrate with shallow open channels and traps to precisely control droplet trajectories. One approach to fabricating these structures involved morphing glass capillary tubes (1.5 mm outer diameter) into various shapes and anchoring them to crosslinked PDMS films, which were bonded to glass slides using an adhesion-promoting primer. We first cleaned the glass capillaries using either oxygen plasma or flame treatment, both of which yielded equivalent results. After cleaning, we functionalized the capillaries by dipping them into the same adhesion primer dissolved in chloroform, then removing them and air-drying. We then placed these primed capillaries onto the cured PDMS film and cast a few drops of uncured Sylgard 184 over them—just enough to wet their surfaces. During this process, the liquid Sylgard wicked between the capillaries and the underlying elastomeric film via capillary action. After 30 minutes, the samples were heated in an oven at 100 °C for one hour, resulting in robust adhesion between the capillaries and the elastomeric film. This bond was highly stable, preventing the capillaries from detaching even when the underlying film was swollen in 5 cSt silicone oil.

Figure 2 (see also Movie S1) shows an example structure produced this way: three bent capillary tubes attached to a PDMS film serve as traps. After swelling the elastomer in 5 cSt silicone oil, we steered a 5 µl droplet of colored water placed on the PDMS film across the substrate. We did so by rotating the supporting stage clockwise and counterclockwise in two orthogonal directions, controlled by an electromagnetically actuated tilt controller (see Figure 1A). As shown in Figure 2, the colored water droplet could be easily moved from one trap to another. Furthermore, reversing the operation could return a droplet captured in the third trap to the first capillary (not shown here).

Notably, when a water droplet contacts a trap, it tends to follow the curved contour by adhering to the bent capillary. This behavior results from capillary interactions between the oil collected beneath the drop and the capillary. Similar behavior was observed with other straight or curved structures, as shown in Supporting Supplementary Materials Figure SM2 and Movie S2.

As the research progressed, we continued to use commercially available motorized biaxial goniometers that could be programmed to run at different speeds and controlled via a joystick.

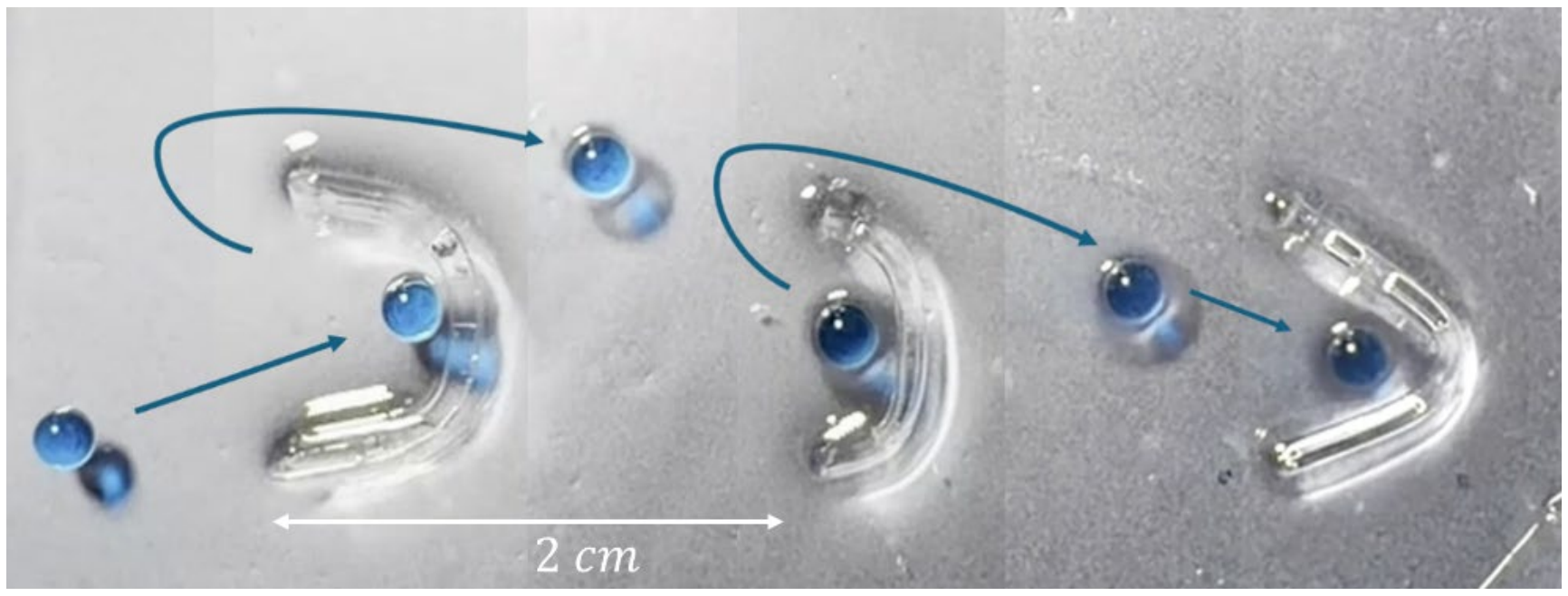

**Figure 2.** Three bent capillaries (outside diameter 1.5 mm) attached to a PDMS substrate serve as traps. The elastomer was swollen in 5 cSt silicone oil. A 5-μL colored water droplet was captured in the first trap on the left and then transported to the other traps by biaxially tilting the substrate. Arrows indicate the direction of droplet motion. This composite image was assembled from multiple video frames.

**Droplets on a Plano-Concave PDMS: Various Scenarios**

Consider a plano-concave PDMS substrate placed symmetrically on a flat plane, with its longitudinal meridian lines emanating from the center and its concentric latitudinal lines converging toward the point of lowest gravitational potential energy, designated as point a. A water droplet released anywhere on this surface will naturally migrate along a longitudinal meridian toward this minimum (Figure 3). If the substrate is then tilted, the center of lowest potential energy shifts to a new position, point b. Consequently, a droplet resting at *a* will move toward *b*. This behavior holds for any substrate orientation, as the droplet consistently follows the shifting center of the latitudinal lines. Thus, by dynamically controlling the tilt of the concave PDMS substrate along two axes, a droplet can be guided along any desired path across its surface. Although the geodesic path (*ab*)—the path of least action followed by the droplet—is technically curved, it is imperceptibly different from a straight line. This is because the substrate's radius of curvature (150 mm) is orders of magnitude larger than the droplet's translation distance (a few millimeters to centimeters). When we placed a droplet at a higher elevation on the plano-concave surface, it moved and decelerated toward its lowest position along a curved geodesic. By tracking the droplet's position over time from a recorded video, we estimated its velocity as a function of its angular displacement about the center of curvature of the PDMS lens (Figure 4). A plano-concave PDMS substrate with a sufficiently low curvature (~0.07 $cm^{-1}$) bonded to a glass slide was ideal for most of the studies described here. Three silicone fluids with viscosities of 5, 10, and 20 cSt were used to swell the crosslinked PDMS. Water drops of various sizes (2 to 5 µl) moved easily on the swelled PDMS substrates. A study of a 5 µl water droplet shows that its velocity increases with the local surface slope and decreases with silicone oil viscosity, as expected (Figure 4). The current understanding of water drop dynamics on an inclined, oil-coated surface has been advanced beautifully by several authors[25,27-31]. When a droplet touches the oil film, a small amount of oil accumulates at its base. Keiser et al.[30,31] summarized the distinct regions of viscous dissipation within the oil film as it moves alongside the droplet, a concept Naga et al.[29] further elaborated through detailed experiments and theoretical modeling. The primary insight from these developments relevant to our current experiments is that the viscous drag force ($F_d$) experienced by a droplet scales with the capillary number ($Ca$) as $F_d \sim \gamma' \, R Ca^{2/3}$ , where $\gamma'$ is the average surface tension of the water and oil, and $R$ is the radius of the droplet. The capillary number is expressed in terms of the droplet velocity ($V$), as well as the surface tension ($\gamma$) and viscosity ($\eta$) of the oil, as $Ca = \eta V / \gamma$. At steady state, this drag force balances the gravitational force acting on the combined mass of the droplet and the oil film accumulated underneath it. Because the density and surface tension of the three oils are nearly identical—with only their viscosities differing—the water droplet's mass remains constant. Consequently, the droplet velocity at any angle of inclination should be inversely proportional to the liquid's viscosity. Furthermore, for a given viscosity, the velocity depends on the inclination angle as $V \sim \theta^{3/2}$.

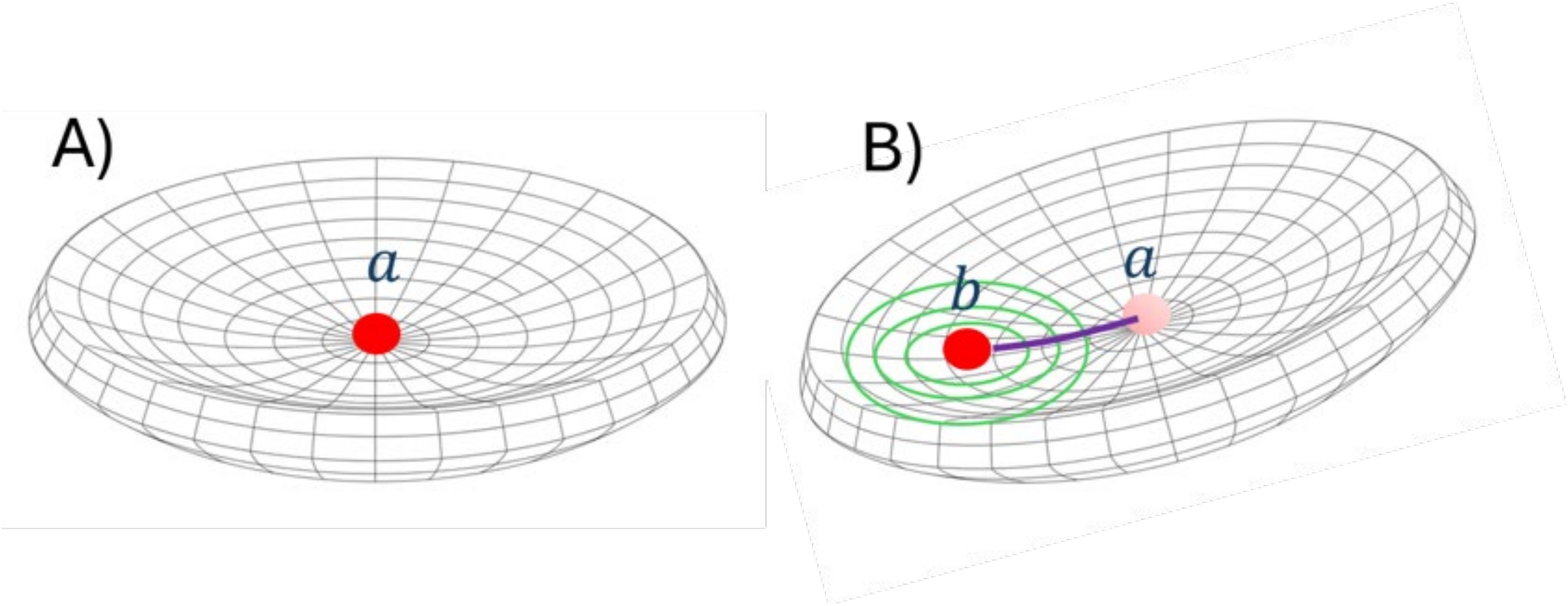

**Figure 3.** The meridian lines on a plano-concave PDMS, where a point of minimum gravitational potential energy always exists at any orientation. A drop always migrates to this point along a geodesic from any other position on the surface.

Notably, these predictions apply to a droplet moving over a nanostructured (or fibrillar) PDMS substrate soaked with oil. The volume of oil accumulated under the droplet is primarily determined by the balance between the increase in gravitational free energy from the rising oil and the consequent decrease in net interfacial free energy. Under this assumption, the oil accumulated under the droplet remains constant during movement, provided that this accumulation minimizes the total free energy. In our system, an additional oil layer exists on the swollen PDMS under gravity. When the substrate is inclined, the oil film thickness increases from the upper to the lower region of the surface. This thickness gradient induces a Laplace force gradient[23] as $F_L =$ $\gamma' \nabla h (\frac{bR^2}{h^2})$, $\nabla h$ being the gradient of the thickness ($h$) of the oil layer, and b is the width of contact. Since, $\nabla h$ is proportional to the local slope of the oil layer, the velocity would scale as $V \sim \theta$. More complex scenarios would arise if we account for non-trivial Landau–Levich type drag[27] and consider both Laplace and gravitational forces. A detailed theoretical analysis of this problem considering the above factors is beyond the scope of the current work. We postpone it for future work.

Given the narrow range of inclination angles and the use of liquids with only three distinct viscosities, we are cautious about using our data to definitively test either theory. We can only conclude that the normalized velocity ($V^* = V \frac{\eta}{\eta_5}$, where $\eta_5$ is the viscosity of the 5 cSt silicone oil) increases qualitatively with the angle of inclination.

These data can guide the selection of an appropriate swelling liquid for specific applications. When using the device for droplet fusion, high viscosity (e.g. $\geq\ 20\ cSt$) silicone fluids slow droplet motion and delay fusion. Conversely, low-viscosity fluids fail to protect droplets from evaporation in high temperature applications because of their low boiling points. Balancing these trade-offs with empirical observations, we determined that swelling the PDMS elastomer in 10 cSt silicone fluid provides an optimal balance for most, including high temperature, applications. While we carried out most of the studies with PDMS swollen in10 cSt silicone fluid, we have also carried out some of the studies with 5 cSt silicone fluid. This distinction is mentioned in the experimental descriptions below.

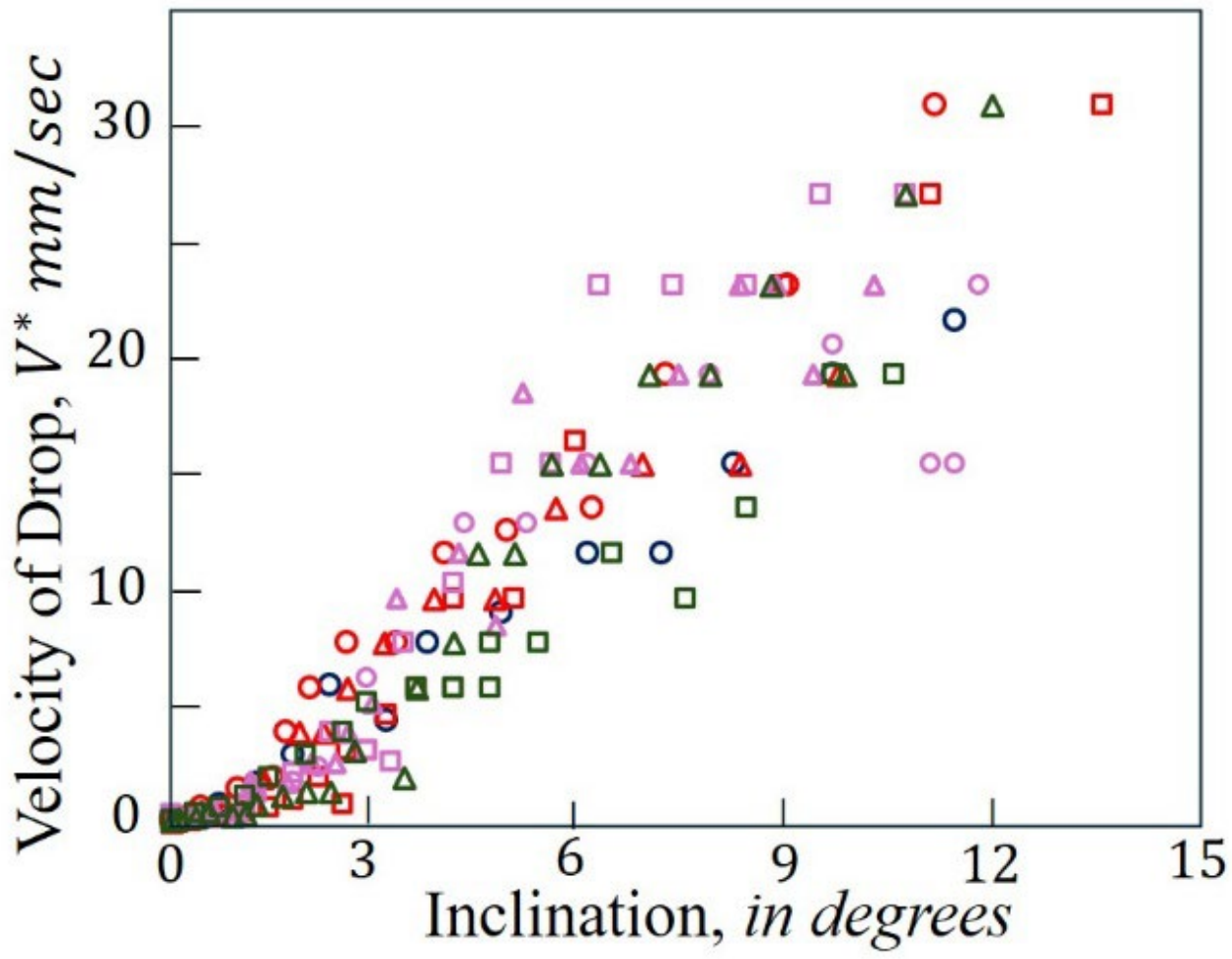

**Figure 4.** The velocity of a 5 $\mu l$ water drop increases with the local slope of a plano-concave PDMS. These experiments used silicone oils with kinematic viscosities of 5 cSt (red), 10 cSt (green), and 20 cSt (pink). Three data sets for each oil are shown. A normalized velocity $V^* = V \frac{\eta}{\eta_5}$ is plotted against $\theta$ .

**Droplet Fusion**

Because a droplet always migrates to the point of lowest potential energy on a plano-concave PDMS surface, regardless of its initial position, this setup is highly advantageous for droplet fusion experiments. Figure 5A illustrates this process: multiple 2 μl water droplets of dilute 0.01 M sodium hydroxide were released onto a plano-concave PDMS substrate containing a 0.01 M nitric acid droplet with an acid-base indicator. The smaller droplets moved efficiently toward the central droplet and fused with it. This acid-base neutralization gradually changed the central droplet's color from purple to green (see also movie S3). A wire mesh placed between the backlight source and the sample enhanced contrast, revealing the oil accumulating around the droplets.

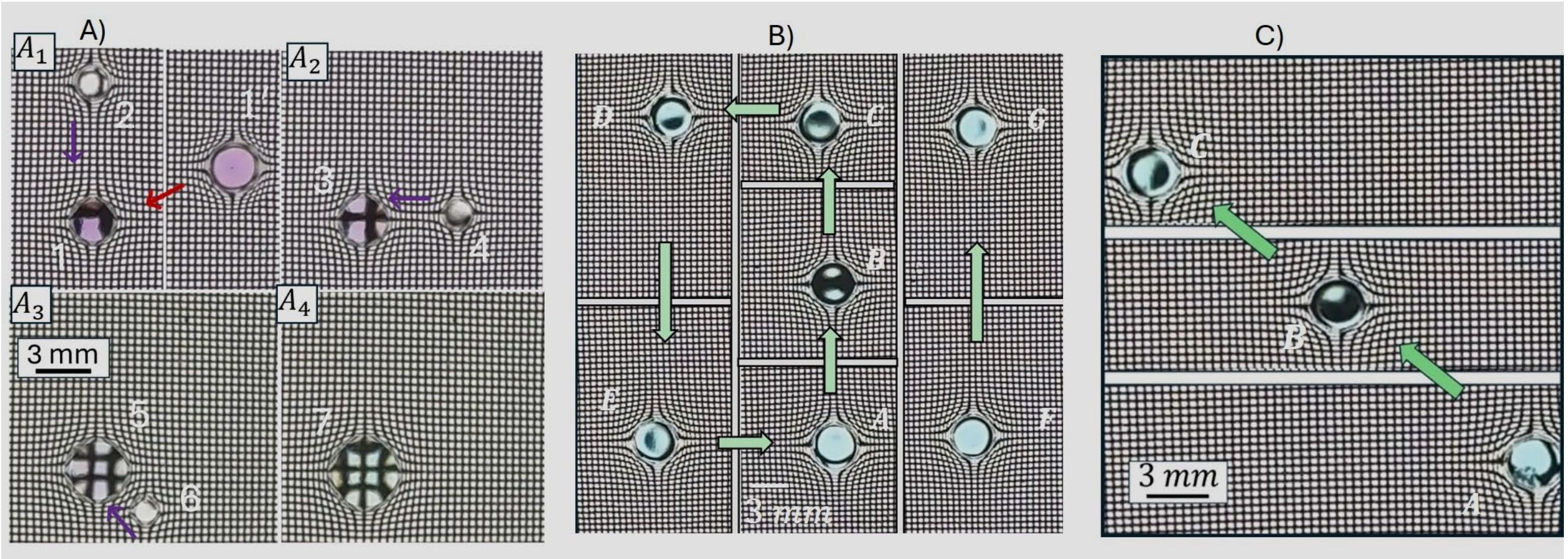

**Figure 5**. A) Here, a 5 $\mu l$ drop of 0.01N $HNO_3$ being titrated by sequential addition of 2 $\mu l$ drops of aqueous NaOH solution (0.01N). B) Directional motion of a 5 $\mu l$ water drop on a plano-concave PDMS, illustrating uniaxial and biaxial motions of a droplet via stepwise counterclockwise rotation ($2^0$ at a time) of the biaxial goniometer. C) Diagonal motion of a droplet is achieved by rotating the two goniometers counterclockwise simultaneously. In all the experiments above, PDMS was swollen in 10 cSt silicone oil.

At this juncture, we highlight an important distinction between the puddle of oil surrounding a droplet at the center of the latitudinal meridian lines (Figure 3B, point b) and the absence of such a puddle (i.e., point a on the tilted substrate). Even when a known amount of oil is added to the plano-concave PDMS to cover its surface, or when excess oil is drained off before a water droplet is applied, oil continues to flow along the surface and accumulate at the base. At equilibrium, this accumulated oil surrounds the water droplet, and part of it moves alongside it if the substrate is tilted (Supporting Supplementary Materials, Movie S4).

Conversely, a droplet that does not start from this equilibrium position accumulates oil from a film that is still draining; hence, the diameter of the oil surrounding this droplet is smaller than that of a droplet starting from equilibrium. Nevertheless, as these two droplets approach each other, their respective oil puddles merge, generating an attractive capillary force that drives the droplets to coalesce. This capillary-driven motion and the subsequent fusion of the droplets are clearly discernible in the video (Supporting Supplementary Materials, Movie S3). Driven by these capillary forces, the center of mass of the fused droplet then shifts toward the region of minimum potential energy—a movement ultimately biased by gravity.

**Directional Motion**

Two additional examples in Figure 5 demonstrate the plano-concave PDMS substrate's versatility in controlling droplet translocation. Figure 5B illustrates an experiment in which one biaxial goniometer was rotated counterclockwise in 2° increments, driving the droplet linearly from position A to C. At the final position, a 2° counterclockwise tilt of the orthogonal goniometer induced a slight lateral displacement. The droplet was then returned to its original position by reversing this sequence (Supporting Supplementary Materials: Movie S5). Furthermore, simultaneously adjusting the tilt angles of the concave PDMS surface in both orthogonal directions allowed the droplet to move diagonally (Figure 5C).

These experiments demonstrate that a droplet can be reliably and efficiently translocated across a plano-concave PDMS surface along multiple geodesics. Although no additional structural modifications are required to guide this motion, such features can be engineered onto the surface to provide greater control.

Based on these results, we conclude that a plano-concave PDMS elastomer coated with silicone oil is an excellent substrate for generating directed droplet motion via biaxial tilting. Furthermore, this method is well-suited for conducting temperature-sensitive chemical reactions (*vide infra*).

We next address how this droplet manipulation method can be scaled to multiple droplets, which requires a mechanism that enables independent "Stop-Go" motion for each droplet. Below, we first describe how this control is achieved with a flat PDMS film, then integrate it with a curved, plano-concave PDMS substrate.

**Asymmetric Motion of Water Drop**

Figure 6 illustrates the fabrication process for a surface that enables unidirectional water droplet transport via confined, inhomogeneous swelling. When an elastomeric polydimethylsiloxane (PDMS) film bonded to a rigid glass substrate is immersed in silicone oil, lateral confinement forces uniform vertical swelling. However, embedding thin, rigid inclusions (such as glass or silicon plates) restricts swelling directly above the plates while allowing normal expansion in the surrounding regions. This inhomogeneous swelling generates a sigmoidal structure on the PDMS surface (Figure 6D), producing a distinct step height. Consequently, a water droplet approaching the sigmoidal region from the right easily descends the step to the left by releasing potential energy (Figure 6E). Conversely, a droplet approaching from the left is pinned by the step, allowing the sigmoidal gate to function as a fluidic diode. A smaller surface step can also be engineered by partially overlapping two embedded plates (Figure 6C). By strategically positioning these steps across the film, the movement of multiple droplets can be precisely controlled, as discussed below (*vide infra*).

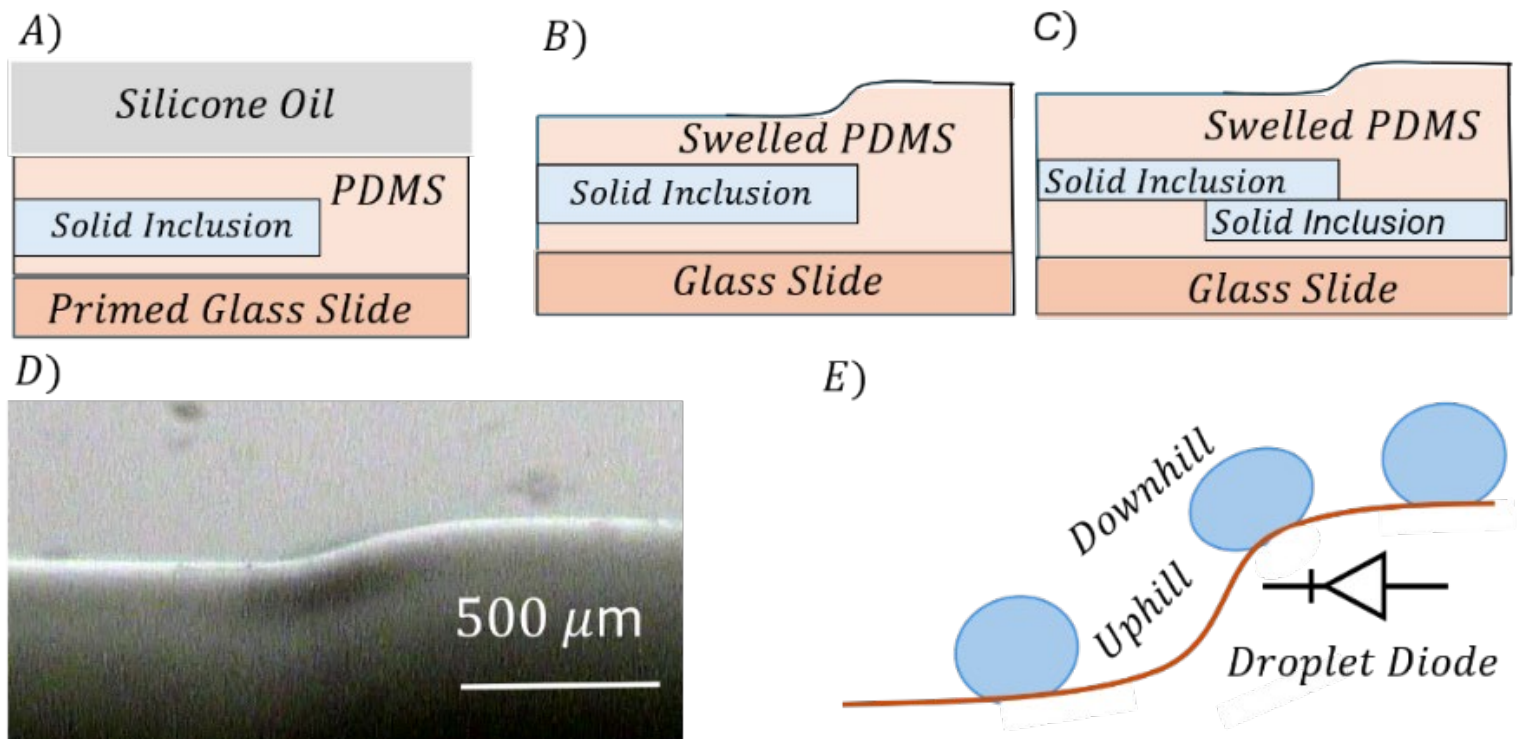

**Figure 6.** Self-explanatory schematic of the substrate preparation for achieving unidirectional motion. Figure D) shows a sigmoidal step created on PDMS using the method described in Figure B). See also the Supplementary Materials: Figure SM3.

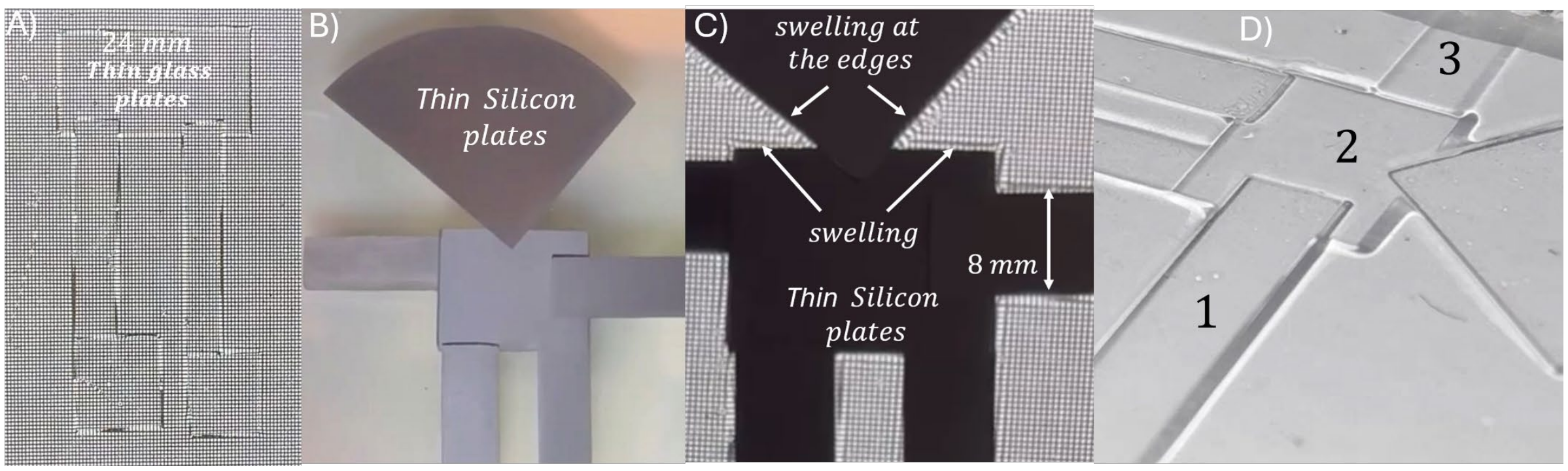

**Figure 7.** Topographies created by inserting thin glass (A) or silicone plates (B) into PDMS and swelling it in 10 cSt silicone fluid. Figure C was taken by passing light through the sample. Figure D corresponds to a PDMS mold of a gelatin imprint of the sample shown in Fig. B.

Figure 7 shows several examples of distinct topographies generated by differential swelling of PDMS in silicone fluid. When thin glass plates are used as the substrate, no surface structures are visible after the elastomer crosslinks. However, once the sample swells in silicone oil, the topography becomes evident when a wire net is placed between the sample and the backlight. In contrast, when a silicon plate is used, the swelling-induced topography is discernible only along the plate edges. A more effective method for examining these surface features is to replicate the swollen PDMS surface using a two-step process (see Experimental Section). First, the original surface is replicated in a gelatin gel, and then a secondary replica of the gel surface is cast using PDMS (Figure 7D).

From the final replica, we can estimate the step height by slightly bending the imprint and imaging the bent region with a telemicroscope. These images, shown in Figures 6 and in the Supplementary Materials SM3, illustrate a sigmoidal variation in the step profile. Based on this profile, the step size is approximately 120 to 200 μm. A more accurate measurement of the step height can be obtained using a Foton probe (Supplementary Materials: Figure SM4).

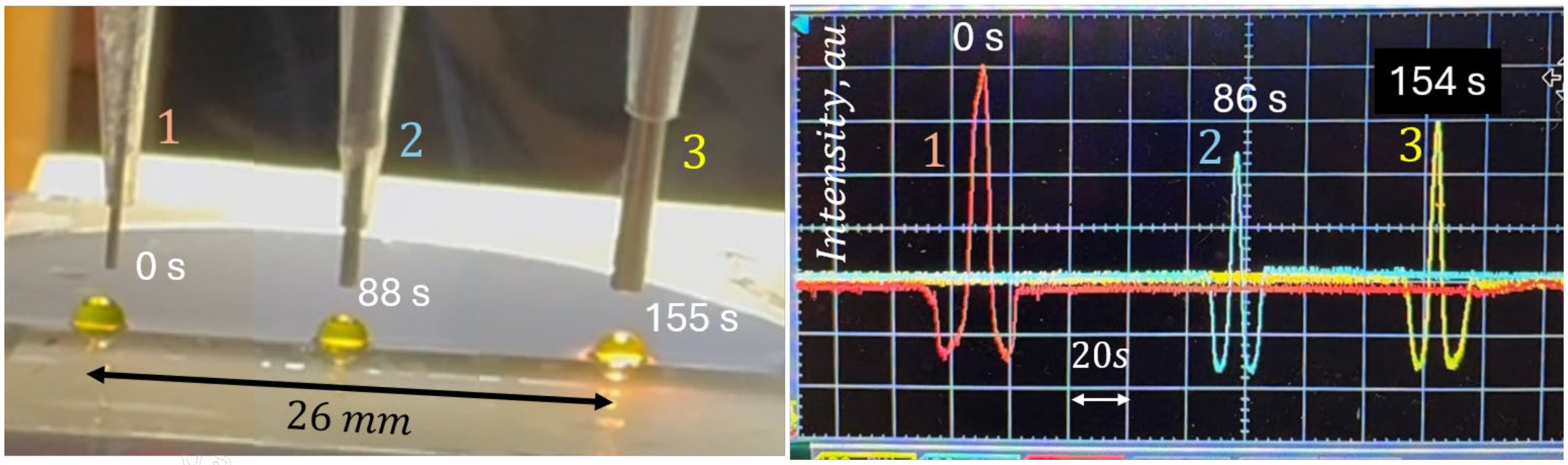

**Figure 8**. Left: A 5 $\mu$l colored water drop moving on Si/PDMS swelled in 20 cSt silicone oil at an inclination of $3^0$. We prepared this image by trimming the original videos into three parts so that, in each segment, the droplet reaches exactly below a Foton probe. Then we combined the three segments into one. Right: Foton spectra of the drop as it travels under the three probes. We set the time the drop passes under the first probe to 0 s; the smallest division on the x-axis is 4s. The probes are numbered as 1, 2, and 3 in color in both figures.

Before describing water droplet motion on these gated structures, it is instructive to review the key characteristics of droplet behavior observed on such surfaces. When a water droplet is placed on the PDMS film, it tends to adhere to the long edge of the swollen region (Figure 8). The Foton probes were precisely aligned above this line.

This clinging effect results from capillary attraction, which arises from the superposition of two concave oil menisci—one beneath the droplet and the other along the edge. When the substrate was inclined at 3°, the droplet moved steadily toward lower potential energy while clinging to the long, straight edge. In these and subsequent experiments, droplet motion was also tracked by one or more Foton probes positioned so that the droplets passed directly beneath them (as indicated by the reflections of the light beams on the top surface of the droplets). The Foton probes captured the droplet's characteristic spectra; these data, combined with the known distance between the probes, enabled us to estimate the droplet's speed. For long-distance, slow motion like the one described here, the timescales from video imaging and Foton spectra are highly comparable. The peaks in the spectrum correspond to instances when the light beam emerging from the probe is perfectly perpendicular to the apex of the droplet. We will discuss the overall shape of the spectrum in a later section.

We investigated the motion of droplets of different sizes on oil-coated PDMS films with various morphologies. For instance, we arranged rectangular silicon or glass strips to partially overlap at their edges, then covered them with liquid PDMS and crosslinked. We then positioned an additional strip to overlap the existing strips at their edges. This setup enabled us to analyze droplet dynamics not only along the long contact lines but also at the short edges formed at the junctions between the silicon strips.

## Motion of Water Drops on the Gated Structures

Here, we summarize the results of several studies tracking the motion of a 5 μL water droplet under three conditions: first, along a long edge at a minimal inclination angle ($\pm 1^0$); second, along the same edge at a higher inclination angle $(5°)$ and third, across a short edge at an $5°$inclination. We conducted these experiments using the sample shown in Figure 9A, which contains thin silicon strip inclusions embedded in PDMS.

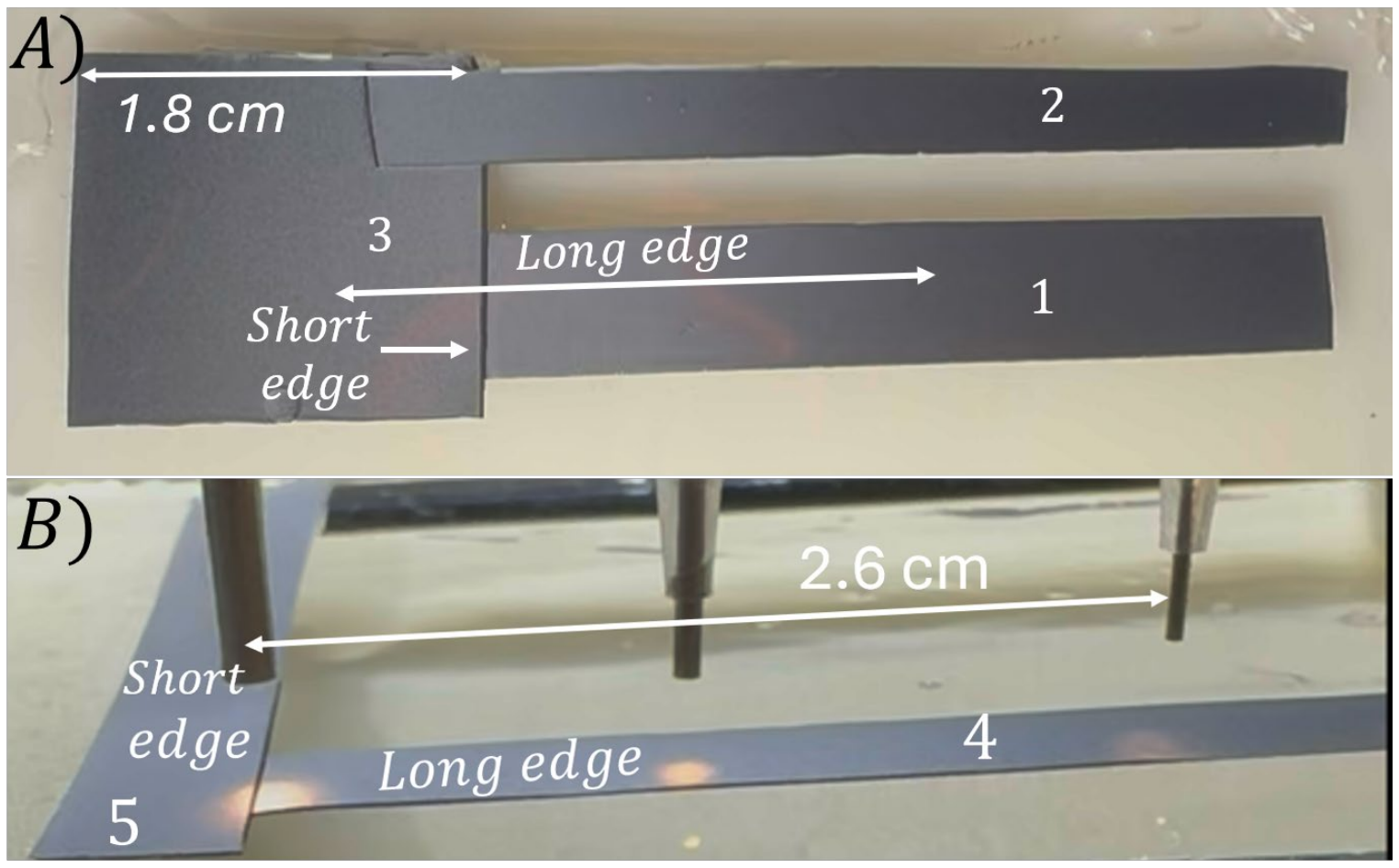

**Figure 9.** Thin silicon strips as inclusions in PDMS. In Figure A, one edge of strip 1 is below 3, whereas the edge of strip 2 is above 3. In Figure B, the edge of strip 4 is below strip 5. When the silicone cures over these strips and swells in 10 cSt silicone liquid, the elastomer will be thicker on 1 than on 5. Similarly, the thickness will be greater on 4 than on 5. Thus, the diode gate developed near the junction will allow a water droplet to move from 1 to 3 or from 4 to 5. A water droplet traveling from 2 to 3 will be arrested near their junction. In Figure B, three Foton probes will detect the water droplet as it passes below them. Note: we could use either the unpolished or the mirror polished side of the silicon wafer. We used the unpolished side as the light was diffusively reflected from its surface, which made the visual observations easy.

Due to the arrangement of these strips, a droplet traveling from arm 1 to station 3 is expected to jump across the short edge. Conversely, a droplet moving from arm 2 toward station 3 should become pinned above the junction of arms 2 and 3. In the present studies, the droplet was placed on the long edge of the front side of strip 1. We recorded droplet movements using a single Foton probe positioned above the sample, supplemented by two microscopes: a telemicroscope to capture the main motion and a second microscope providing an orthogonal view. This setup, combined with precise hydraulic manipulation and motion controls, ensured that the Foton probe remained strictly perpendicular to the droplet's upper surface throughout the experiments.

**Fotonic Sensor Used to Study Droplet Profile**. A Fotonic sensor[26] consists of two sets of optical fiber bundles; light is transmitted to the sample through one bundle, while the diffusively reflected light is collected by the other. When the probe is perpendicular to a surface, most of the reflected light reaches the receiving bundle, yielding maximum intensity. In our setup, this maximum occurs when the probe is aligned perpendicularly to the droplet's apex. Otherwise, the light beam reflects off-specularly from the curved surface, so the probe primarily collects low-intensity, diffusely scattered light from the moving droplet.

Specifically, when a concave oil meniscus surrounds the base of the droplet, diffuse scattering begins at the meniscus as the droplet enters the optical field. Subsequently, light is diffusely reflected from the droplet's convex surface. Starting at the meniscus edge, the light intensity recorded by the Fotonic probe decreases and remains low until the probe passes a critical point on the droplet's surface (Figure 10). At this critical point, the tangent to the droplet's surface forms an angle close to 45°. The probe then captures more diffusely reflected light until the intensity peaks at the apex of the droplet. As the droplet continues to move past the central axis, the intensity decreases, producing a profile that is nearly a mirror image of the initial trajectory. While precisely converting this optical intensity to extract the exact profiles of the droplet and its surrounding oil meniscus is outside the scope of the present work, we intend to pursue this analysis in the future.

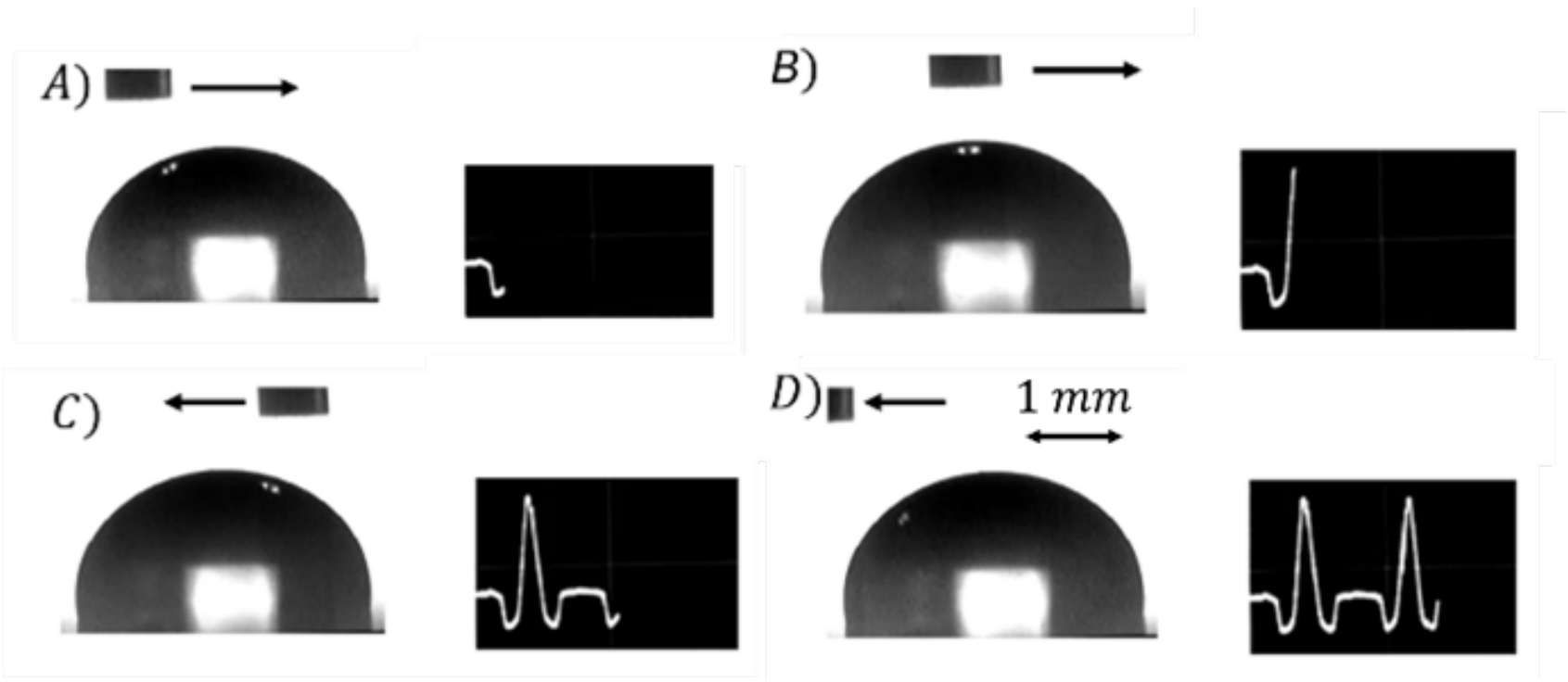

**Figure 10.** This figure shows video snapshots of a Fotonic probe moving at a steady pace over a glycerol. It illustrates how the intensity of diffusely reflected light from the droplet varies with the probe's position above the droplet. We chose glycerol because it does not evaporate, allowing ample time to conduct various experiments. See also Supplementary Materials: Figure SM5 and Movie S6)

**$1^0$ Inclination, Test of Reversibility**

We now describe droplet motion along the long edge of a substrate inclined at only 1°. As noted above, we captured droplet motion using both video microscopy and a Fotonic probe. The optical profile yields several important qualitative insights. First, the region with a more gently varying slope (Figure 11) corresponds to the length of the oil meniscus surrounding the droplet. Second, by comparing the symmetry of the photon spectra profiles as the drop moves to the right or left under the probe, we can assess the reversibility of the motion. Any discrepancies between these profiles would indicate a probe misalignment, necessitating adjustments to the substrate tilt. The Fotonic spectra shown in Figure 11B, captured after these adjustments, indicate that a reversible motion of the drop has been achieved at $\pm 1^0$ inclination, although it is not perfect. The difference in the overall shape of the blue- and red-colored spectra indicates that the drop moves slightly faster in one direction than the other. With laborious, time-consuming adjustments, these spectra can be made to overlap, but this is not very practical when multiple measurements need to be made.

### Uniform Drop Motion at 5⁰ Inclination

We next examine droplet motion at a higher inclination angle on the same substrate (Figures 11C and 11D). In these experiments, we dispensed 5 µl water droplets onto the surface using a minipump with a 25-gauge needle at a rate of one droplet every 12 seconds. We deposited the droplets gently near one end of the long edge and slid them downward along the substrate, which was inclined at 5°.

A striking difference between this motion and that observed at a 1° inclination is the smaller volume of oil accumulating around the droplet; furthermore, oil depletion occurs after only a few droplets pass. Despite this depletion, subsequent droplets continue to move sequentially along the edge in a train-like fashion at the same 12-second intervals. Foton probe spectra collected for several consecutive droplets (Figure 11E) are remarkably consistent. Notably, the near-plateau region (Figure 11B) evident during droplet motion at a 1° inclination is absent here.

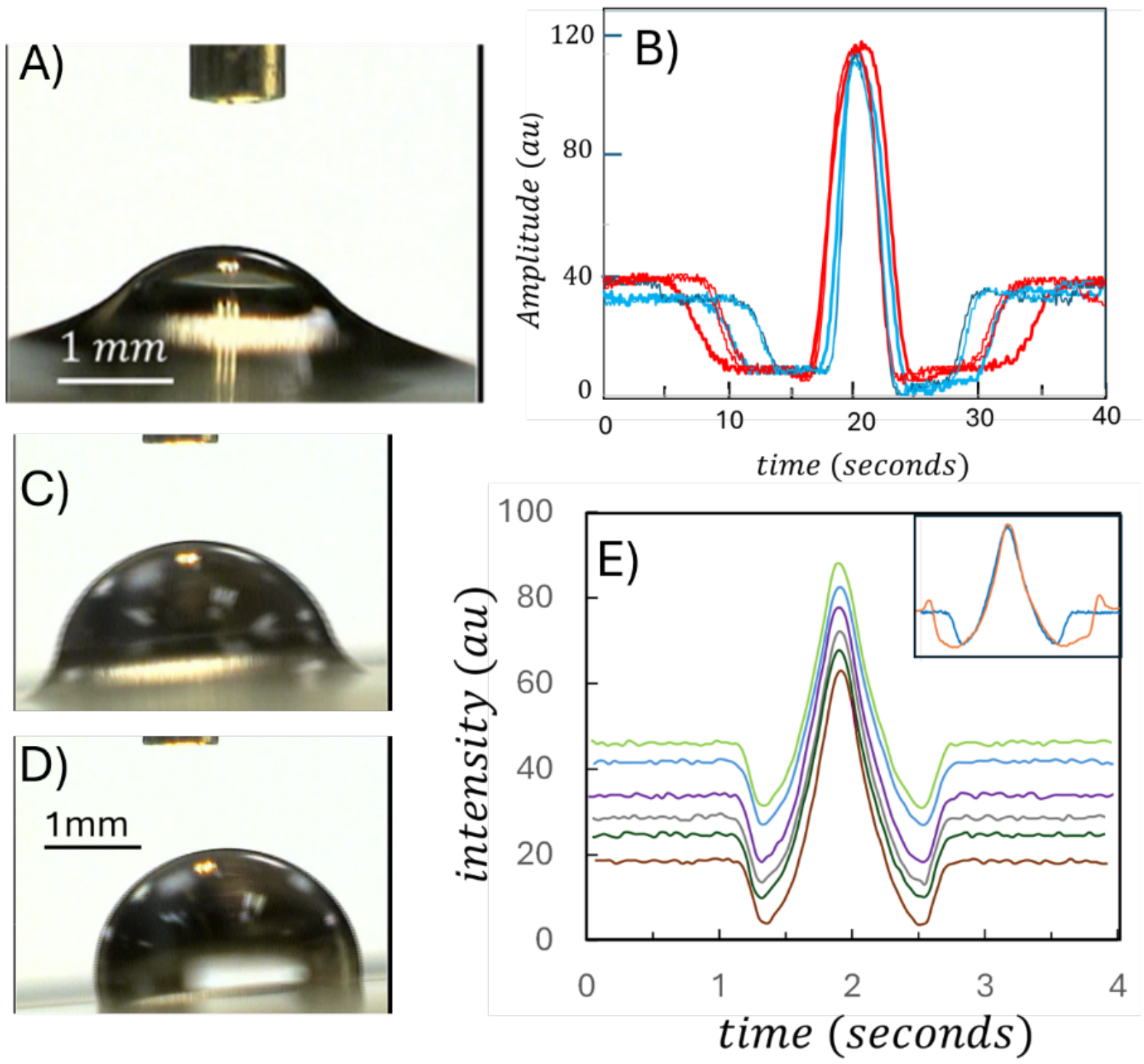

**Figure 11.** (A) A colored 5 µl water droplet moving on a substrate at a ±1° inclination. (B) Fotonic spectra of the droplet moving at a ±1° inclination; the red and blue curves correspond to movement from left to right and right to left, respectively. (C, D) Front views of a water droplet moving on the substrate at a 5° inclination. (E) Superimposed Fotonic spectra of several droplets. The inset compares the Fotonic spectra of a droplet moving over a small accumulation of underlying oil (C, red line) versus after this oil layer has been depleted (D, blue line).

**Motion over the Sigmoidal Gate at the Short Edge**

The video microscopic images and Fotonic spectra of droplets passing over an edge at a 5° inclination exhibit characteristics similar to those described in Figure 11E; notably, no large pauses are observed. Nevertheless, several snapshots from a single recorded video show that the droplet moves relatively uniformly during the first 0.5 seconds before accelerating abruptly over the next 0.25 seconds (Figure 12A). The Fotonic spectra (Figure 12B) capture remarkably similar behavior; all droplets initially move slightly slower without coming to a complete halt, and then they run off speedily in a highly consistent manner. Because every droplet accelerates past a specific point (indicated by the dotted circle in Figure 12E), we conclude that this behavior implies the existence of a weak energy barrier that impedes motion but does not completely arrest it.

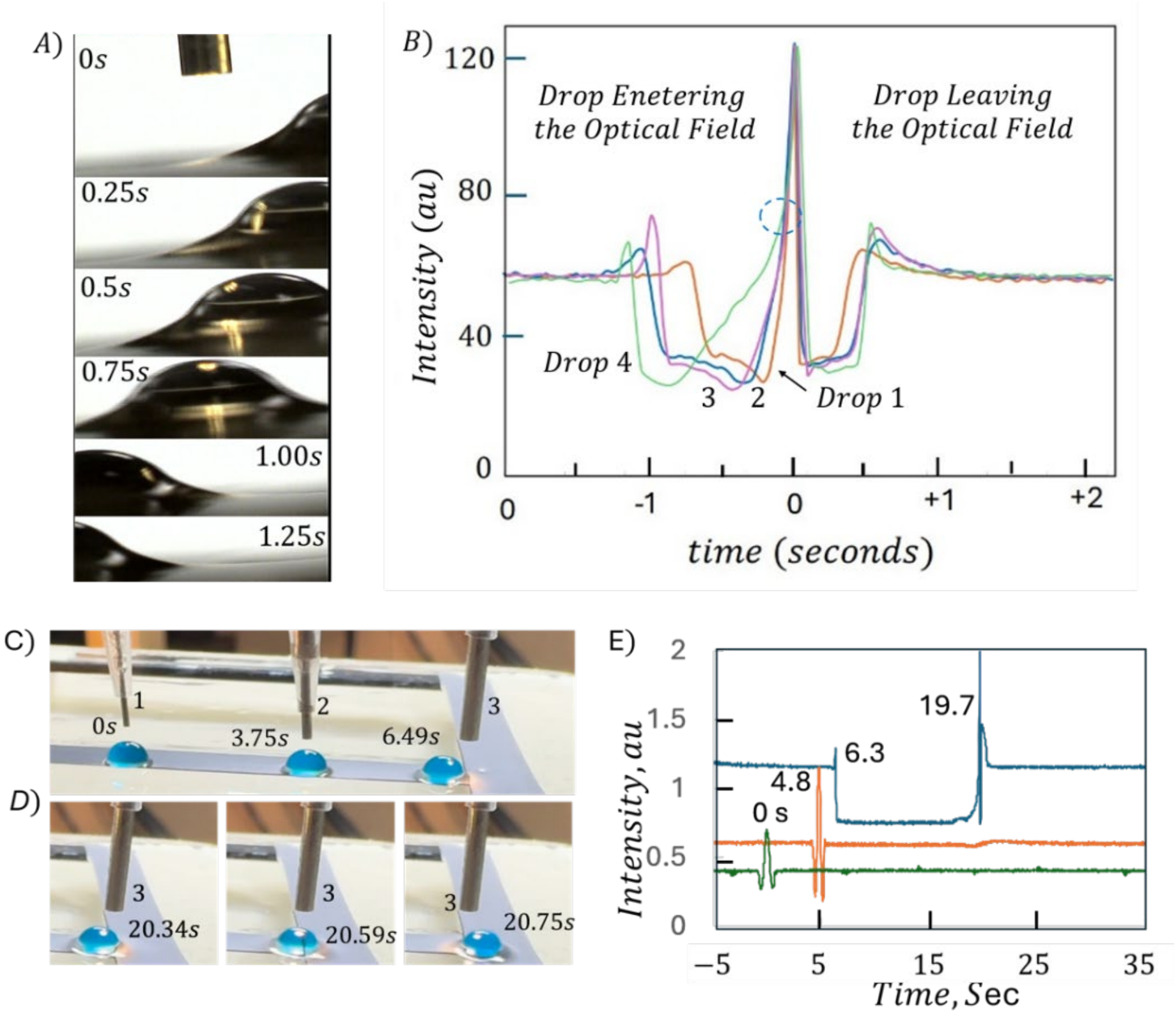

**Figure 12**. Optical observation in A and Fotonic spectra in B illustrate that a 5 µl droplet moves on a diode-like gate in an unstable manner; it first slows down and then suddenly speeds up after a critical point is reached (indicated by the dotted circle). Optical (C and D) and Fotonic probe observations (E) correspond to a 10 µl colored water droplet moving at a 4° inclination on a silicone-oil-impregnated PDMS structure (Figure 9B) that was swollen in 5 cSt silicone fluid.

We now discuss another experiment in which a 10 µl colored water droplet moves at a 4° inclination on the substrate shown in Figure 9B. In this experiment, the three Foton probes—positioned slightly above the substrate along the long edge—detect the droplet as it passes beneath them. A key observation from this experiment, consistent with the one described in Figure 8, is that the droplet clings to one of the long edges while translating along a silicon strip (Figure 12C), However, an interesting dynamic arises when the droplet encounters a short edge perpendicular to its path, as shown beneath Foton probe 3 in Figures 12C to 12E.

Upon reaching this boundary, the droplet pauses briefly before rapidly crossing the edge (Supporting Supplementary Materials: Movie S7). This pause duration increases with subsequent droplets, a consequence of the gradual depletion of the oil accumulated around the droplet's base. In one specific instance involving a late-sequence droplet, a remarkably long pause was recorded. However, when a droplet of oil was deposited onto the substrate behind the stalled droplet after 14 seconds, the water droplet accelerated and crossed the edge almost immediately (Figure 12D). These sequential events of deceleration and subsequent acceleration are a weak reminder of the single-droplet dynamics summarized in Figure 12B, although their time scales are very different.

At this juncture, we highlight an interesting study recently published by Wu et al.[22], which investigated the gravity-induced motion of glycerol droplets on a soft PDMS substrate with a pre-fabricated crease. As the droplets moved across the ridge perpendicular to the crease line, three distinct behaviors were observed: large droplets crossed the crease, medium-sized droplets became trapped within it, and small droplets stopped just before reaching it. This final observation is qualitatively similar to the behavior reported here.

In many physical systems, the interplay between a non-linear energy landscape and a linear driving potential gives rise to stable and metastable energy states that govern various phenomena, including contact line pinning. In the system described by Wu et al[22], the coupling of the elastocapillary energy with the gravitational potential energy creates these stable and metastable states.

While the insight of Wu et al. is novel and effectively explains their observations, the elastocapillary length of our elastomer (~0.01 μm) is considerably smaller than theirs (~20 μm). This stark contrast suggests that the droplet pinning observed in our system may stem from a different mechanism. If we consider only the long pause shown in Figures 12C and 12D, we might envision that the pinning is caused by the surface-tension force exerted on the water droplet by the concave oil film behind it. Nevertheless, the unstable motion observed in Figure 12A requires a different explanation, as it occurs mid-journey. Initially, we hypothesized that the droplet encounters a local topographic barrier, or "bump," whose height is small enough to decelerate the droplet rather than pin it entirely. However, we could not definitively confirm the presence of such surface topography.

Conversely, another observation prompts us to reconsider this scenario. When the excess oil from the sample swollen in 10 cSt silicone oil (Fig. 9A) was removed with detergent and distilled water, subsequent deposition of a few water droplets revealed crater-like deformations—roughly 1.5 mm in diameter—visible to the naked eye (Supplementary Materials, Figure SM6). Because a swollen elastomer can behave differently from an unswollen one, this finding suggests that a mechanism similar to the one proposed by Wu et al. cannot be entirely ruled out. We plan to investigate this behavior in depth in future work.

**Motion of Droplets Over Asymmetric Gates On a Plano Concave Platform**

The studies outlined above lay the groundwork for the complex manipulation of a single droplet. However, a significant challenge in designing functional droplet devices is the need to manipulate multiple droplets simultaneously in different directions over numerous cycles with high reliability. The studies reported above clearly demonstrate that controlling the accumulation and depletion of oil under a water drop is critical. A plano-concave PDMS platform is advantageous because its concavity preserves the total amount of oil by allowing it to redistribute across the surface. A planar surface can also conserve oil, provided it is not used repeatedly for motion in a single direction. In a typical fluidic operation, the substrate is expected to undergo clockwise and counterclockwise tilts along its two orthogonal axes. Because several surface channels also trap oil, the overall redistribution of oil is not a significant concern.

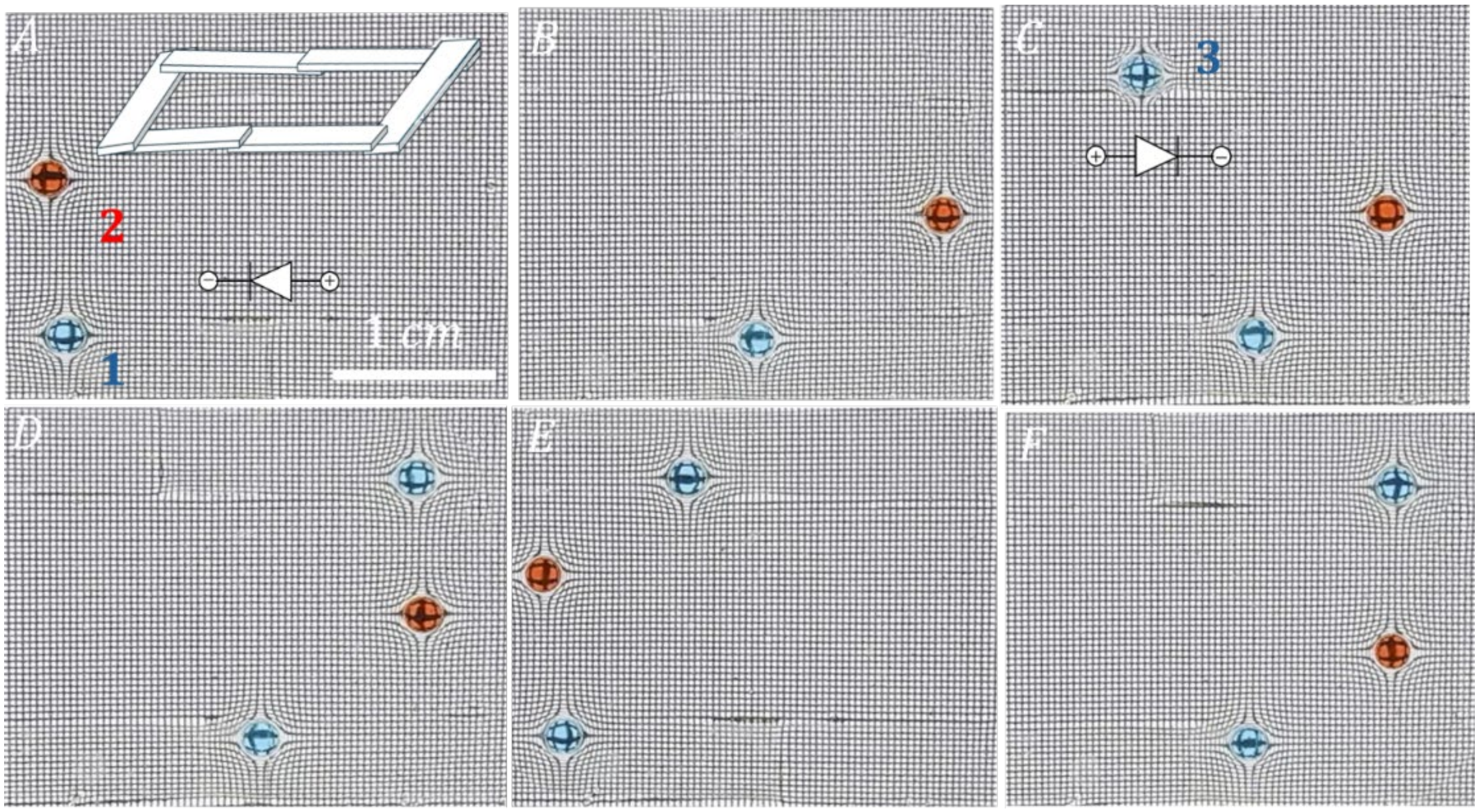

**Figure 13.** Figure A corresponds to an inclination of the concave PDMS as -4$^0$; Blue drop (1) and Red drop (2) are moving to the right. The orientation of diode is shown. Figure B corresponds to an inclination is +5$^0$. Fig C, blue drop (3) is released, orientation of diode is shown. Figure D: blue drop 3 and the red drop 2 are parallel, Fig E corresponds to an inclination of -4$^0$, Figure F corresponds to an inclination of +5$^0$. The radius of curvature of the plano concave PDMS was 150 mm, while its diameter in the plan view was 59.6 mm.

Here, we report progress in manipulating multiple droplets using both plano-concave and flat PDMS substrates. We first introduce a method for manipulating multiple droplets using a plano-concave PDMS substrate. This substrate is embedded with thin rectangular glass plates arranged as shown in the inset of Figure 13A. Because the plates partially overlap, crosslinking of the elastomer and its subsequent swelling in silicone oil create several fluidic diodes on the surface. These diodes allow droplet motion in one direction while arresting it in the opposite direction (Figure 13 and Supporting Materials: Movie S8). Consequently, when a droplet is placed on the lower arm of the structure and the substrate is tilted slightly clockwise, the droplet moves along the bottom edge. It then stops when the substrate orientation is altered by just one or two degrees at a time.

In contrast, a red droplet placed at the center of the substrate experiences no such restriction. Consequently, when the entire substrate is rotated clockwise by 5°, the blue droplet becomes trapped upon reaching a fluidic gate, whereas the red droplet, which is in the middle of the substrate, continues to move forward. If another blue droplet is introduced behind the first two at the other arm, it continues moving until it aligns parallel to the red droplet (Figure 13D). When the substrate is subsequently rotated counterclockwise, both the red droplet and the first blue droplet move to the left; however, as the second blue droplet reaches the fluidic diode gate, it becomes trapped. This process can be repeated cyclically. Furthermore, a single droplet can be driven clockwise in a closed loop due to the specific design of the fluidic diodes, while counterclockwise movement is strictly prevented.

More complex architectures can be fabricated by embedding multiple, variably arranged rigid plates within the PDMS. These interconnected networks allow droplets to be selectively translocated from one channel to another, depending on the desired functional application of the device.

**Diode Controlled Manipulation of Multiple Droplets**

Next, we demonstrate the manipulation of multiple droplets based on these principles, using a flat PDMS film with an embedded intricate structure. At the center of this structure is a rectangular station made from cut pieces of a silicon wafer. Four rectangular strips are arranged around it: two partially overlap the top of the central station, while the other two overlap it from below. Additionally, a semicircular wafer overlaps the central station, with its pointed end positioned above the station.

A replica of this structure in another PDMS elastomer, prepared by imprint lithography, is shown in Figure 14A. This method ensures that the boundary between the long Si strip and the background PDMS is sigmoidal (see Figure 6D). While estimating the step heights at the junctions of the two PDMS strips is generally difficult, precise measurements can be made using a pre-calibrated Foton probe translated slowly over the step. In Supplementary Materials: Figure SM4, the reflection of light from the optical probe onto the PDMS surface is clearly visible, allowing unambiguous identification of the PDMS–air interface.

The Foton probe is calibrated by positioning it at the interface and moving it vertically upward. At each vertical increment, the distance between the probe and the surface is measured with a microscope, establishing a calibration curve for the Foton probe readings. Once calibrated, the probe is translated laterally to the higher region of the step. The difference between the two Foton probe readings is then converted to step height using the calibration. When this PDMS film was swollen in 10 cSt silicone oil, the step height along the long edge was estimated to be approximately 250 μm. A similar analysis of the step formed by the partial overlap of two silicon plates yielded a step height of about 90 μm.

The directions of the fluid diodes are illustrated in the replica shown in Figure 14A (see also Supporting Supplementary Materials: Movie S9). According to this design, when the substrate is inclined clockwise, a droplet placed in region B will move to the central station by crossing the short edge separating regions B and C. Conversely, a droplet placed in region D will be arrested by the short edge separating regions D and C, even if the substrate is rotated in the orthogonal clockwise direction. On the other hand, droplets are free to move from E to C or from G to C. All of these events occur at a substrate tilt angle of ≤5°. A larger tilt is required only to move a droplet against a diode gate arrest. When three or four droplets fuse, they could pass through this arrest at a tilt of about 5°.

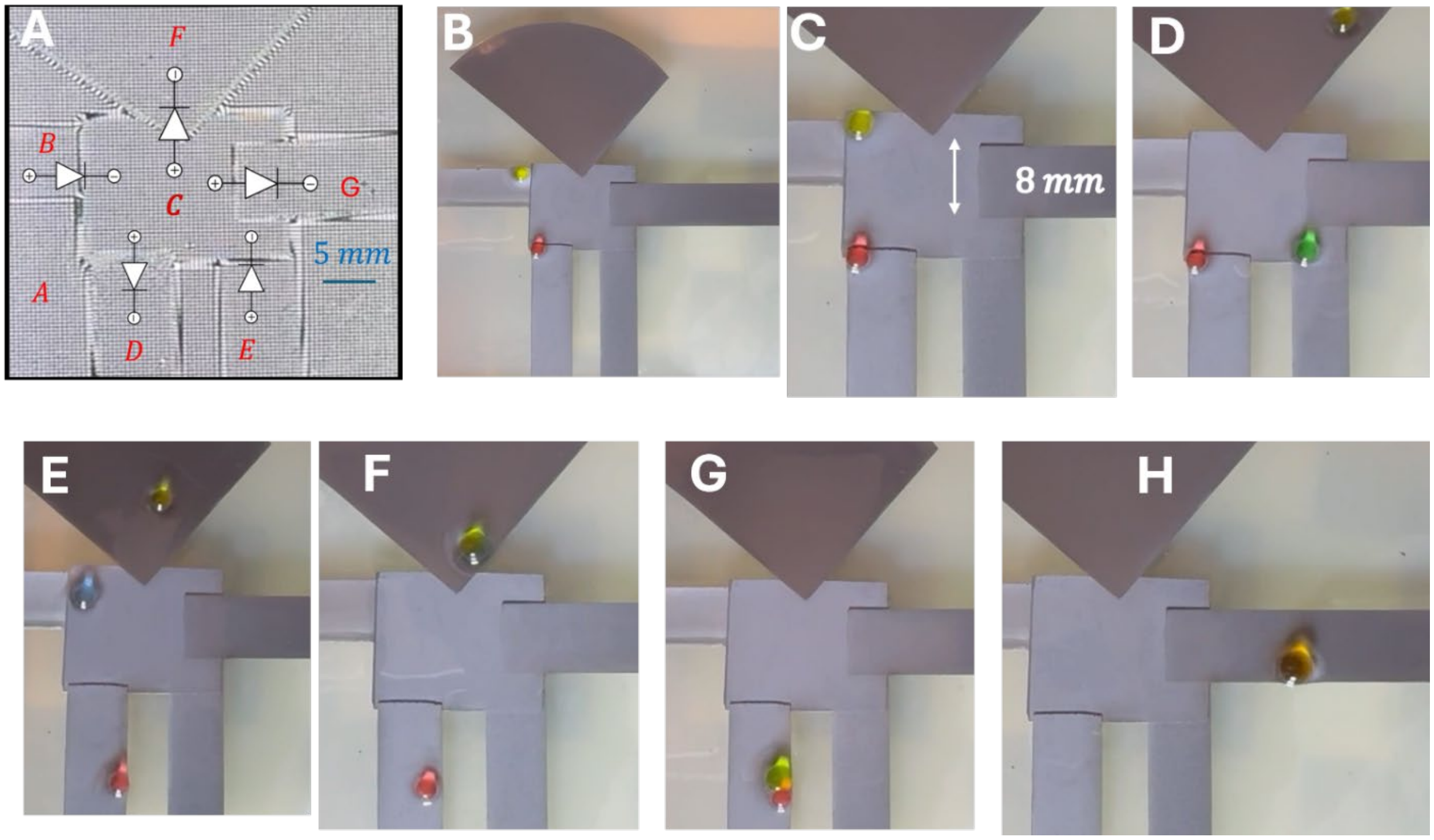

**Figure 14**. Figure A shows a PDMS mold taken from a gelatin imprint of the sample (featured in Figure 6), captured with a backlight and a wire net between the light source and the sample. The expected directions of motion for a water droplet are indicated by diode symbols. Figures B and C show that the yellow droplet successfully crosses the boundary between sectors B and C. In contrast, the red droplet is arrested at the edge between sectors C and D. Figures B through H display the sequential movement of various colored droplets, including their fusion and subsequent migration to other sectors of the sample. The tilt angles in all these experiments varied from $2^0$ to $5^0$.

**Example of a Thermally Activated Reaction**

Whether a microfluidic device will have practical applications depends largely on its ability to support thermally activated chemical reactions. In this study, we demonstrate that these PDMS platforms are suitable for studying temperature-dependent reactions. As a proof of concept, we performed the well-known redox process used for glucose analysis, in which glucose reacts with Benedict's reagent at approximately 100 °C on a plano-concave PDMS substrate.

Heating of the PDMS platform was achieved by attaching a thin-film heater to the backside of the glass support and passing a 0.6 A current through it. A FLIR camera was used to map the surface temperature of the plano-concave PDMS. When a drop of an aqueous D-glucose solution (0.17 wt% in water) and a drop of Benedict's reagent were deposited at two different locations on the PDMS, they moved toward each other and coalesced (Figure 15). The slightly bluish, fused droplet initially settled into a 42 °C zone on the surface. In the FLIR camera's color-mapping mode, the droplet appeared dark blue at this stage. As the concave PDMS was tilted toward the warmer side using a joystick-controlled biaxial goniometer, the droplet traversed a 72 °C region, where it remained slightly bluish to the eye, while appearing light aqua blue according to the thermal camera's color code. Upon further tilting, the droplet reached the center of the heater's warmest zone. At this point, the droplet visually turned translucent and then brown, while the thermal camera recorded its color signature as red.

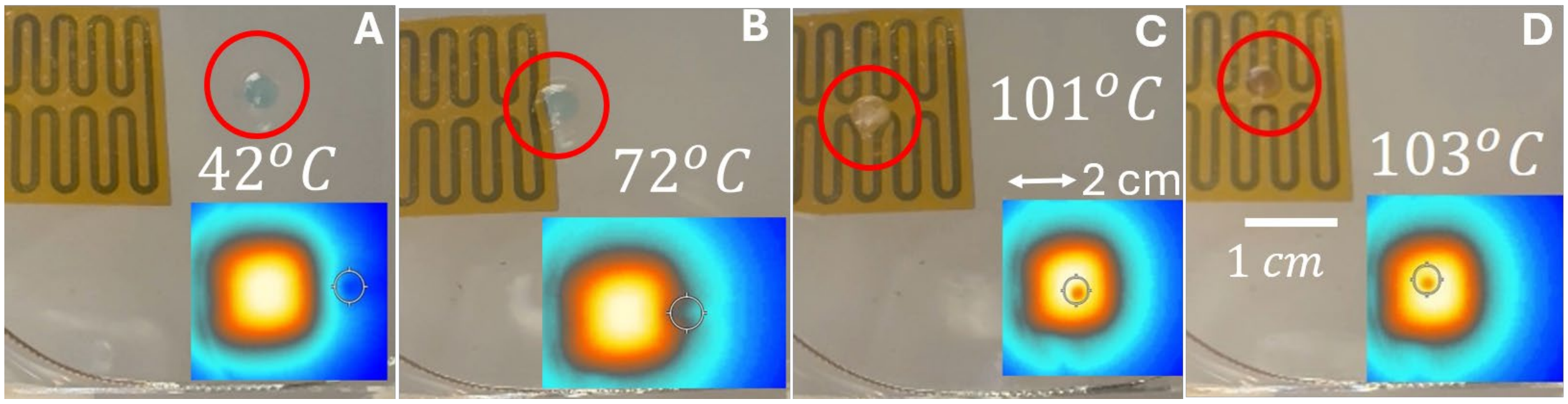

**Figure 15.** Droplet motion on a plano-concave PDMS substrate coated with a thin layer of 10 cSt silicone oil. The photographs in Figure A were captured 3 seconds after the sucrose solution and Benedict's reagent droplets merged. Following this fusion, the resulting 10 µl droplet remained at that position (42 °C) for approximately 10 seconds before moving toward the edge of the thermal element (72 °C). After resting at the 72 °C zone for one minute, the droplet entered the heated region. Within another 30 seconds, it was moved toward the center of the heated zone, where its temperature rose to 101 °C (Figure C). The droplet was then moved across the surface for about three minutes to ensure uniform heating, after which it turned brown (Figure D).

An important point to note is that as the droplet[1] moves from a colder to a hotter region of the substrate, the thermal Marangoni force generated within both the oil film and the droplet opposes this motion. This resistance is particularly pronounced when the droplet attempts to cross the edge of the heating film (e.g., Figure 15B, see also Supporting Supplementary Materials: Movie S10). We observed similar behavior in our previous study[17] (see the caption of Figure 11 in this reference), where substrate vibration helped overcome Marangoni-force-induced droplet pinning).

## Consolidation of Important Points

What we present here demonstrates that the biaxial tilt of a PDMS substrate swollen with silicone oil is suitable for controlling the motion of water droplets on its surface. This approach is closely related to our earlier report[23], where local tilts were generated on a PDMS film via localized deformations. In both cases, the driving force is governed by the combined actions of gravity and a Laplace pressure gradient. While we focused here mainly on the results obtained with PDMS, other substrates can also produce decent results. In fact, Elias and Fruge[24] conducted their initial studies using polypropylene, which has shown reasonable success. The main point to emphasize is that the droplet should be repelled by the underlying substrate, which can be predicted on the basis of their dielectric properties using Lifshitz theory[28,32] , or simply in terms of surface and interfacial tensions [23,33] ensuring that the free energy of adhesion is positive, thus preventing the oil film from thinning out completely during the transit of the droplet.

At this juncture, we wish to comment on the swelling of PDMS by silicone oil—a well-established subject that has been recently reviewed[34,35]. In our own previous work[36,37] on this topic, we found that the slip and release properties of oil-loaded PDMS depend not only on macroscopic variables, but also on material-specific parameters, such as the oil's molecular weight and the elastomer's elastic properties. In the current study, we used a commercial PDMS (Sylgard 184) and a silicone oil. Moving forward, we aim to conduct a more detailed investigation to systematically optimize these compositions. Such studies would be particularly crucial for soft elastomers that exhibit elastocapillary deformations.

The movement of a single drop on a tilted plano-concave PDMS substrate, as described here, shows that the droplet follows a geodesic path along a longitudinal meridian. This path is bounded by higher-energy meridian lines, which effectively act as a surrogate channel. Under slow steering, the droplet remains in undifferentiated equilibrium; consequently, it stops almost immediately if the external drive is removed, preventing droplet runaway. Together, these characteristics are ideal not only for droplet translation but also for droplet fusion and thermally activated chemical reactions. To provide a proof of concept, we presented an example requiring only a single heating element. However, it is possible to integrate multiple thermal zones on the concave PDMS to carry out more complex reactions that require multi-level heating and cooling, such as a polymerase chain reaction (PCR). In the future, we will discuss how to further improve the heating mechanism itself, as well as how to isolate different thermal zones using appropriate thermal barriers on PDMS.

## Conclusions

Our research introduces a method for routing multiple fluid droplets on PDMS substrates using joystick-controlled biaxial tilts. We combine a silicone oil-swollen flat or plano-concave PDMS substrate with sigmoidal physical steps that function as one-way diodes, guiding droplets along predictable paths. The experimental results show that these designs can efficiently achieve steering, translocation, and fusion of droplets.

## Acknowledgements

This research was supported by the Franklin J. Howes Jr. Distinguished Professorship, which the senior author held until retiring on December 31, 2025. The research began in the summer of 2025. Parts of this research were presented at the AICHE meeting in 2025. MKC presented and discussed some of the work described here with Prof. Howard Stone's group at Princeton University early this year. The excellent discussions with members of the Stone group inspired some of the studies described here. During manuscript preparation, we used Google Gemini and Grammarly to polish the English and check grammar in the text we wrote. We generated the sketches shown in Figure 1B and Figure 3 interactively using Adobe Firefly. The original photograph used to sketch Figure 1B is shown in the Supplementary Materials: Figure SM1. We carefully reviewed and edited the entire document to ensure accuracy. The authors take full responsibility for the final content, scientific validity, and integrity of the published work.

**References**

1. Link, D. R.; Anna, S. L.; Weitz, D. A.; Stone, H. A. Geometrically Mediated Breakup of Drops in Microfluidic Devices. *Phys. Rev. Lett.* **2004**, *92* (5), 054503.
2. Utada, A. S.; Lorenceau, E.; Link, D. R.; Kaplan, P. D.; Stone, H. A.; Weitz, D. A. Monodisperse Double Emulsions Generated from a Microcapillary Device. *Science* **2005**, *308* (5721), 537–541.
3. Chaudhury, M. K.; Chakrabarti, A.; Daniel, S. Generation of motion of drops with interfacial contact. *Langmuir* **2015**, *31* (34), 9266–9281.
4. Huo, J.; Gou, X.; Zhang, J.; Zhu, J.; Chen, F. A review of droplet/bubble transportation on bionic superwetting surface. *Small* **2025**, *21* (19), No. 2412363.
5. Miao, J.; Zhang, Y.; Wang, L.; Tsang, A. C. Bioinspired Adaptive Surfaces for Intelligent Liquid Manipulation: Progressing From Passive and Active to Hybrid Strategies. *Adv. Mater.* **2026**, DOI: 10.1002/adma.202674360.
6. Zhang, J.; He, S.; Wang, H.; Bai, L.; Sun, M.; Huang, J.; Wang, D.; Lai, Y. Vapor-mediated micro/nanodroplet manipulation on superhydrophilic surface for bioassay analysis. *Chem. Eng. Sci.* **2026**, 123966.
7. Dai, H.; Gao, C.; Sun, J.; Li, C.; Li, N.; Wu, L.; Dong, Z.; Jiang, L. Controllable high-speed electrostatic manipulation of water droplets on a superhydrophobic surface. *Adv. Mater.* **2019**, *31* (43), No. 1905449.
8. Xie, M.; Qian, Z.; Wang, X.; Li, Y.; Shuai, Y.; Wang, Z.; Wang, Z. Bionic structured milli-fluidics: a review. *Chem. Rev.* **2026**, *126* (2), 1347–1407.
9. Bai, H.; Zhao, T.; Cao, M. Interfacial fluid manipulation with bioinspired strategies: special wettability and asymmetric structures. *Chem. Soc. Rev.* **2025**, *54* (4), 1733–1784.
10. Jahangir, R.; Nasirimarekani, V. Magnetic Droplet Manipulation on Open Surfaces. *Adv. Mater. Technol.* **2026**, DOI: 10.1002/admt.202671007.
11. Tan, J.; Du, J.; Lv, D.; Gao, Y.; Jiang, D.; Wang, Z.; Liu, W.; Tso, C. Y. Orbital Electrowetting: From Continuous Droplet Transport to Programmable Microfluidics. *Adv. Mater.* **2026**, *38* (33), No. e73401.
12. Abe, T.; Oh-hara, S.; Ukita, Y. Integration of deep reinforcement learning to simple microfluidic system toward intelligent control: Demonstration of simultaneous microbeads manipulation. *Sens. Actuators, B* **2023**, *397*, No. 134636.
13. Gyimah, N.; Scheler, O.; Rang, T.; Pardy, T. Deep reinforcement learning-based digital twin for droplet microfluidics control. *Phys. Fluids* **2023**, *35* (8), No. 082001.

14. Guo, K.; Song, Z.; Zhou, J.; Shen, B.; Yan, B.; Gu, Z.; Wang, H. An artificial intelligence-assisted digital microfluidic system for multistate droplet control. *Microsyst. Nanoeng.* **2024**, *10* (1), No. 138.
15. Lagzi, I.; Soh, S.; Wesson, P. J.; Browne, K. P.; Grzybowski, B. A. Maze solving by chemotactic droplets. *J. Am. Chem. Soc.* **2010**, *132* (4), 1198–1199.
16. Mukhopadhyay, A. K.; Niu, R.; Fu, L.; Feng, K.; Fujta, C.; Zhao, Q.; Qu, J.; Liebchen, B. Automated decision-making by chemical echolocation in active droplets. *Proc. Natl. Acad. Sci. U.S.A.* **2026**, *123* (5), No. e2526773123.
17. Daniel, S.; Chaudhury, M. K.; de Gennes, P.-G. Vibration-Actuated Drop Motion on Surfaces for Batch Microfluidic Processes. *Langmuir* **2005**, *21* (9), 4240–4248.
18. Mertaniemi, H.; Jokinen, V.; Sainiemi, L.; Franssila, S.; Marmur, A.; Ikkala, O.; Ras, R. H. Superhydrophobic tracks for low-friction, guided transport of water droplets. *Adv. Mater.* **2011**, *23* (26), 2911–2915.
19. Britto, S.; Kujala, Z.; Lee, S.; Gonella, S. Drops on architected elastic substrates: A repertoire of regimes at the turn of a knob. *Phys. Rev. Res.* **2025**, *7* (3), No. 033095.
20. Charara, M.; Kujala, Z.; Lee, S.; Gonella, S. Spatially selective drop-motion programming using metamaterials. *Proc. R. Soc. A* **2025**, *481* (2308), No. 20240429.
21. Britto, S.; Gonella, S. Centipede-Like Metastrip Enables On-Demand Programmable Droplet Motion. arXiv:2608.20409. Submitted August 20, 2026.
22. Wu, Z.; Linton, G.; Karpitschka, S.; Pandey, A. Creases as elastocapillary gates for autonomous droplet control. *Proc. Natl. Acad. Sci. U.S.A.* **2026**, *123* (39), No. e2600758123.
23. Biswas, S.; Pomeau, Y.; Chaudhury, M. K. New drop fluidics enabled by magnetic-field-mediated elastocapillary transduction. *Langmuir* **2016**, *32* (27), 6860–6870.
24. Elias, V.; Elias, Fruge, M., Gating of Droplets in Open Surface Microfluidics https://aiche.confex.com/aiche/2025/meetingapp.cgi/Paper/723984;
Fruge, M.; Elias, V. Drop Fluidics Enabled By Gravity and the Modulation of Laplace Pressure https://aiche.confex.com/aiche/2025/meetingapp.cgi/Paper/723708
25. Raufaste, C.; Cox, S. J.; Celestini, F. Actuating water droplets on liquid infused surfaces: A rickshaw for droplets. *Phys. Rev. Fluids* **2021**, *6* (8), No. 083603.
26. Hall, A.; Bundy, M. *Overview of Piezoelectric Actuator Displacement Measurements Utilizing an MTI-2100 Fotonic Sensor*; Technical Report ARL-TR-5551; Army Research Laboratory: Adelphi, MD, USA, 2011.
27. Cantat, I. Liquid meniscus friction on a wet plate: Bubbles, lamellae, and foams. *Phys. Fluids* **2013**, *25* (3), No. 031303.
28. Daniel, D.; Timonen, J. V.; Li, R.; Velling, S. J.; Aizenberg, J. Oleoplaning droplets on lubricated surfaces. *Nat. Phys.* **2017**, *13* (10), 1020–1025.
29. Naga, A.; Rennick, M.; Hauer, L.; Wong, W. S.; Sharifi-Aghili, A.; Vollmer, D.; Kusumaatmaja, H. Direct visualization of viscous dissipation and wetting ridge geometry on lubricant-infused surfaces. *Commun. Phys.* **2024**, *7* (1), No. 306.
30. Keiser, A.; Keiser, L.; Clanet, C.; Quéré, D. Drop friction on liquid-infused materials. *Soft Matter* **2017**, *13* (39), 6981–6987.
31. Keiser, A.; Baumli, P.; Vollmer, D.; Quéré, D. Universality of friction laws on liquid-infused materials. *Phys. Rev. Fluids* **2020**, *5* (1), No. 014005.
32. Chaudhury MK; Good RJ Retarded van Der Waals Force Theory of Stable Thick Films in Systems of Three Condensed Phases. J. Colloid Interface Sci. 1983, 94 (1), 292–294

33. Li, C.; Yu, J.; Schehr, J.; Berry, S. M.; Leal, T. A.; Lang, J. M.; Beebe, D. J. Exclusive Liquid Repellency: An Open Multi-Liquid-Phase Technology for Rare Cell Culture and Single Cell Processing. ACS Appl. Mater. Interfaces 2018, 10 (20), 17065.
34. Boucher, D. G.; Safaripour, M.; Croll, A. B.; Webster, D. C. Oil-infused silicone elastomers for barnacle and ice release: The current state of understanding. *Prog. Polym. Sci.* **2025**, *164*, No. 101966.
35. Kolle, S.; Ahanotu, O.; Meeks, A.; Stafslien, S.; Kreder, M.; Vanderwal, L.; Cohen, L.; Baur, J. A.; Weaver, J. C.; Aizenberg, J. On the mechanism of marine fouling-prevention performance of oil-containing silicone elastomers. *Sci. Rep.* **2022**, *12* (1), No. 11799.
36. Vorvolakos, K.; Chaudhury, M. K. The Role of Interfacial Slippage in Adhesive Release. In *Microstructure and Microtribology of Polymer Surfaces*; Tsukruk, V. V., Wahl, K. J., Eds.; ACS Symposium Series 741; American Chemical Society: Washington, DC, 1999; pp 234–248.
37. Chaudhury, M. K.; Vorvolakos, K.; Malotky, D. Friction at soft polymer surface. In *Polymer Thin Films*; Tsui, O. K. C., Russell, T. P., Eds.; World Scientific: Singapore, 2008; pp 195–219.

**Supporting Supplementary Materials**

# Directional Control of Droplet Motion via Geometric Gating

Manoj K. Chaudhury, Vanessa Elias and Michelle Fruge

Chemical and Biomolecular Engineering, Lehigh University
Bethlehem, PA 18015 (USA)

## Experimental Details

## Materials

The materials for the synthesis of *adhesion primer* were as follows: bis (triethoxysilyl) ethane, vinyl trimethoxy-silane, Titanium n-Butoxide (TBT), and platinum-Divinyl Tetramethyl Disiloxane complex in xylene (Karstedt catalyst) (all these chemicals were purchased from Gelest Inc., USA); hydrogen silsesquioxane (HSQ, H-resin, Dow Corning, USA); Chloroform (Carolina Biological Supply, USA). The materials for acid-base titration were: HNO3 (0.01N, EM Sciences) , NaOH (pellets, EM Sciences), Methyl Purple indicator (Rica Chemical Co). 0.2 g of the indicator was dissolved in 7.3 gm 0.01M $HNO_3$.

Silicones: Sylgard-184 (two-part kit, Dow Corning Corp); Silanol (1000 cSt) and trimethoxy silane terminated silicone oils of viscosities 10 and 20 cSt were purchased from Gelest; 5 cSt silicone oil was purchase from Sigma Aldrich. Chloroform (HPLC grade, Sigma-Aldrich); Acetone (99.5 %, Fisher). Glass cover slips (22 mm x 22 mm, 0.2 mm) were purchased from Corning Glass. For glucose analysis: D-Glucose (Fisher); Benedict's reagent (Carolina Science); Adhesive Polyimide Heater Elements Films (2.8W 5V, 25mm x 20mm x 0.3 mm) were purchased from Amazon. Silicon wafers 2" diameter, 280 micron thickness were purchased from Electron microscopy sciences. Glass cover slips (22 mm x 22 mm, 0.2 mm) were purchased from Corning Glass. Clean rectangular (7.5 cm x 5 cm x 1 mm) glass slides were purchased from Electron Microcopy Sciences. The square glass slides (3 in x 3 in, 1.8 mm thick) were purchased from United Scientific. Watch glasses (10 cm diameter) were purchased from Fisher. Biconvex glass lenses were purchased online from eBay. Water used in these studies was deionized (DI, Thermo Scientific Barnstead E-Pure Unit)

## Analytical and Process Instruments

Foton probes (MTI-2100 Fotonic sensor, MTI Instruments, Inc., USA) were used for most of the experiments reported here. For couple of experiments two additional Foon probes (MTI 1000 and MTI KD 370A) were used; Motorized goniometers and Joystick control (Zaber); Manual goniometers (Melles Griot), Three axis motion controllers with Joysticks (Narishighe) and one hydraulic manipulator for the control of photon probe (Narishighe) were purchased from Tritech Research, Inc. An additional Three axis motion controller (Narishighe) was purchased from Nikon. The infrared camera (FLIR C5) was purchased from Edmund Optics.

## Preparation of the primer

The primer was prepared by mixing the following components (given in terms of weight percent in chloroform): HSQ (3.3 %), bis (triethoxysilyl) ethane (0.2 %), vinyl trimethoxy silane (0.9 %), TBT (0.2 %), and platinum catalyst (0.033 %), which produced a clear colorless liquid. The prepared solution (500 ml) was stored in a refrigerator, which increased the shelf life of the solution.

The details can be found in: Al-sakkaf, Khulood, Monicka Kullappan, and Manoj K. Chaudhury. "Adhesive Release via Elasto-Osmotic Stress Driven Surface Instability." *arXiv preprint arXiv:2303.11437* (2023).

**Bonding of PDMS to Glass Slides**

A flame-treated, clean glass slide was rubbed with a Kimwipe moistened with a few drops of the primer solution, then air-dried. Concurrently, Sylgard 184 was prepared by mixing the base elastomer and curing agent at a 10:1 weight ratio, per the manufacturer's instructions. After degasification in a vacuum oven, a few drops (0.06 g/cm2) of the mixture were deposited at the center of the glass slide. The slide was placed on a pre-leveled surface for several hours to overnight to allow the mixture to spread evenly, then cured at 100 °C for one hour. This protocol ensured complete crosslinking and robust adhesion of the elastomer to the glass substrate. Strong bonding was essential to prevent the PDMS coating from delaminating during subsequent swelling in PDMS fluid for periods ranging from days to weeks. When necessary, the surface of the swollen PDMS was cleaned with a dilute soap solution and distilled water, then dried with clean air supplied by an FTIR purge gas generator.

**Imprint Lithography**

A 10 wt% gelatin solution was used to replicate a substrate via imprint lithography. To eliminate air bubbles at the interface between the gelatin and the hydrophobic polydimethylsiloxane (PDMS) substrate, a few drops of a superwetting surfactant (Dow Corning) were added to the solution. The gelatin solution was heated to ~80 °C and sonicated until all air bubbles were removed and the solution became completely transparent.

This clear solution was poured over the substrate in a square polystyrene Petri dish, ensuring the entire sample was covered. Once the gelatin reached room temperature, the Petri dish was placed in a refrigerator to allow the gelatin to set. After 24 hours, the gelatin mold was gently peeled away from the substrate, leaving a highly defined negative replica.

Next, a freshly prepared, clear mixture of Sylgard 184 was poured over the negative gelatin imprint. The setup was transferred to a cooling chamber and kept at ~10 °C for at least one week to ensure sufficient initial crosslinking of the PDMS. To complete crosslinking, the entire sample was then cured at 100 °C. Finally, the crosslinked PDMS was repeatedly rinsed with tap water, followed by distilled water, to remove any residual gelatin.

**Preparation of Plano-Concave and Flat PDMS**

To fabricate the plano-concave PDMS structure, one side of a double-convex glass lens (15 cm radius of curvature) was replicated. The lens was first cleaned with soapy and distilled water, then subjected to flame treatment. Both sides of the lens were coated with silanol-functionalized polydimethylsiloxane (PDMS) and heated in an oven at 100 °C overnight. Unreacted silicone was removed by rinsing multiple times with chloroform, followed by washes with a dilute soap solution and running distilled water. Residual water droplets adhering to the lens were blown off with clean air supplied by an FTIR purge gas generator.

A specific amount (~0.06 g/cm²) of degassed PDMS was placed at the center of a flat glass slide that had been modified with an adhesive primer, as described previously. Immediately afterward, the hydrophobized convex glass lens was gently placed over the PDMS-coated glass on a pre-leveled platform. After leveling for one day, the PDMS was cured in an oven at 100 °C for a couple of hours. Once the assembly cooled to room temperature, the glass lens was separated by gently peeling it from one end after initiating a small crack at the edge of the PDMS–lens interface with a sharp razor blade.

**Study of Droplet Motion**

A plano-concave polydimethylsiloxane (PDMS) substrate was prepared by curing Sylgard 184 against a convex glass lens, as described in the experimental section. The convex template's radius of curvature ranged from approximately 80 mm to 200 mm, but a radius of 150 mm was used for most of the studies reported here.

Initial testing showed that a water droplet did not move freely on this pristine surface because of contact angle hysteresis-induced contact line pinning. To eliminate this hysteresis, the PDMS was swollen in silicone oil with a kinematic viscosity of 5 to 20 cSt. After several days of swelling, the excess oil was decanted, the PDMS dishes were cleaned, and a few fresh drops of silicone oil were deposited onto the surface.

For the majority of the experiments, 5 μl water droplets were deposited onto the surface. The substrate was then tilted clockwise and counterclockwise using a biaxial goniometer, as described below. Under these conditions, the angular displacement of the droplet remained within 10° of the center of the concave lens. To ensure consistency across certain comparative studies, the exact amount of topically applied oil was standardized by uniformly spreading 72 mg of silicone oil over a surface area of approximately 45 cm².

In some experiments, food coloring was added to the water to improve visualization; control tests confirmed that the dye did not affect droplet motion. The experimental setup featured a concave polydimethylsiloxane (PDMS) substrate integrated with a metal wire screen. Droplets were precisely deposited at [Insert specific location, e.g., the center of the concave PDMS surface / the wire screen]. The system demonstrated high sensitivity, with droplets as small as 1 to 2 microliters exhibiting distinct motion. All experiments were conducted using silicone oil with a kinematic viscosity of 10 cSt. The radius of curvature of the plano-concave PDMS was 15 cm, whereas its diameter in the plan view ranged from about 5 cm to about 7 cm.

**Preparation of diode-like gates**

Silicon strips (6–7 mm wide, 33–45 mm long) or glass strips (22 mm × 5 mm) were cut from as-received silicon wafers or glass coverslips. The strips were sonicated in a 50:50 (v/v) mixture of chloroform and acetone, then subjected to a gentle flame or plasma treatment. Next, the strips were coated with the adhesion primer (vide supra) and placed on a glass slide previously coated with a partially crosslinked PDMS film. They were then covered with a thin layer of premixed, de-aerated Sylgard 184 and kept on a pre-leveled stage overnight at room temperature. Finally, the assembly was cured at 100 °C for approximately three hours.

For specific studies, the primed silicon or glass strips were arranged so that their edges partially overlapped. The strips were aligned either linearly or in a perpendicular geometry [Note: Please verify the missing text here, e.g., to yield 10 cSt or 20 cSt films, respectively]. Finally, the samples were allowed to swell in silicone fluids for periods ranging from 24 hours to several weeks.

**Apparatus for Controlled Bi-Axial Tilting**

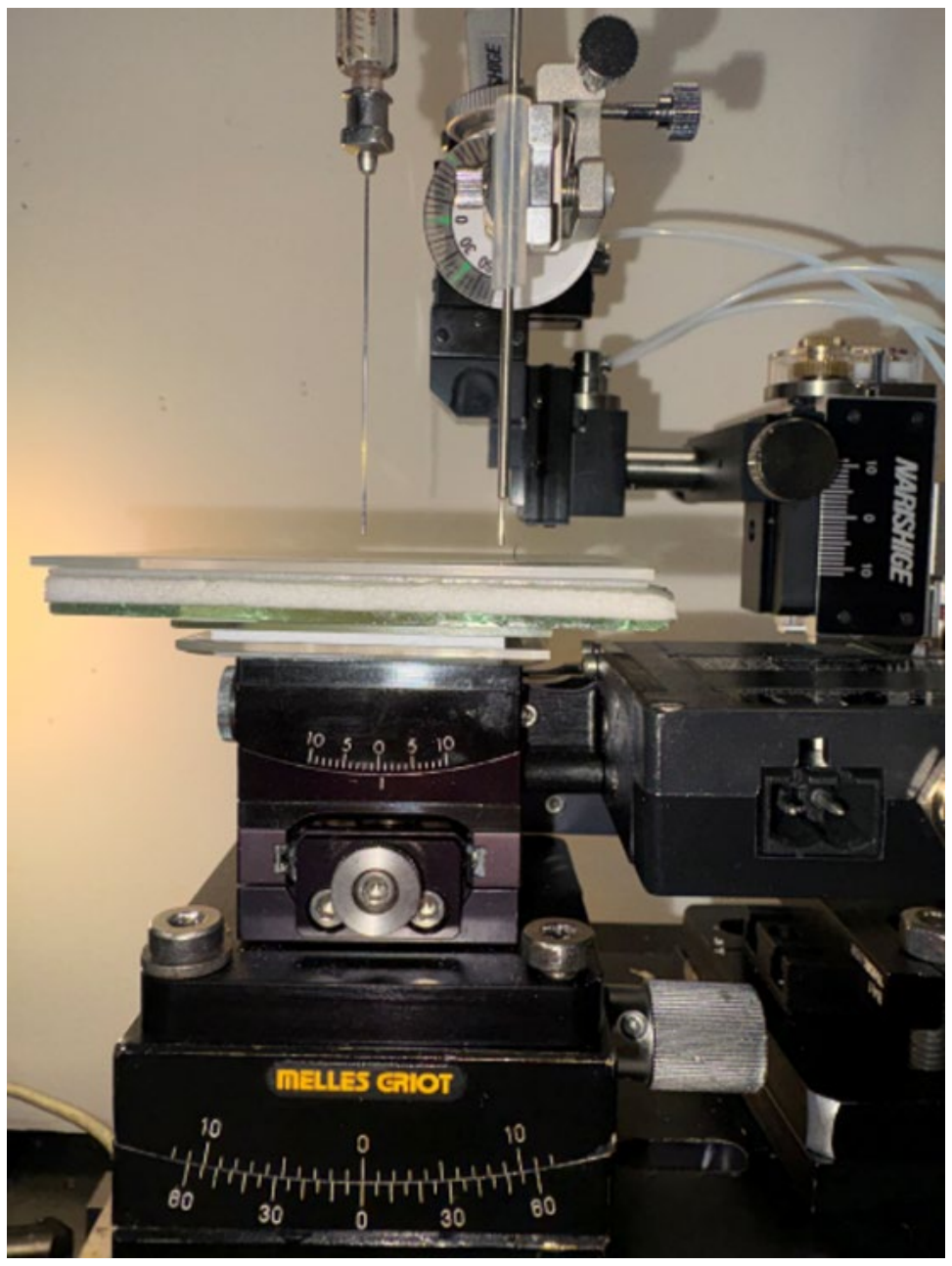

**Figure SM1.** An original photograph of the Motorized bi-axial goniometers, a sketch of which is presented in the Text (Figure 1).

The experimental apparatus used to perform the droplet motion experiments is detailed in the main text (Figure 1). This setup was used to induce droplet motion on swollen PDMS films by mounting the substrates on a stage and tilting it bi-axially with joystick-controlled motorized stages. Droplets were delivered onto the sample surface either via a micropipette (Eppendorf, 1–10 µl range) or through a 25-gauge steel needle. For the latter method, water was dispensed using a minipump (Thomas Scientific) at a flow rate of one 5-µl droplet every 12 seconds, generating a continuous train of droplets that passed beneath the Foton probe.

Droplet dynamics were monitored using two tele-microscopes positioned orthogonally. One microscope recorded the droplets' lateral motion, while the other confirmed that the Foton probe remained precisely centered above the moving droplet. For experiments requiring a wider field of view, an iPhone camera was used to record video footage. These videos were subsequently processed using Adobe (Adobe Express, Adobe Firefly) and Microsoft (e.g., Clipchamp) software suites.

As noted in the text, Elias and Fruge22 previously used a silicone oil-coated PDMS substrate with shallow, open channels and traps to precisely control droplet trajectories. Supplementary details about their experimental setup are provided here: the substrate was tilted in two orthogonal directions using a hand-built, electromagnetically controlled tilting device (Figure SM1).

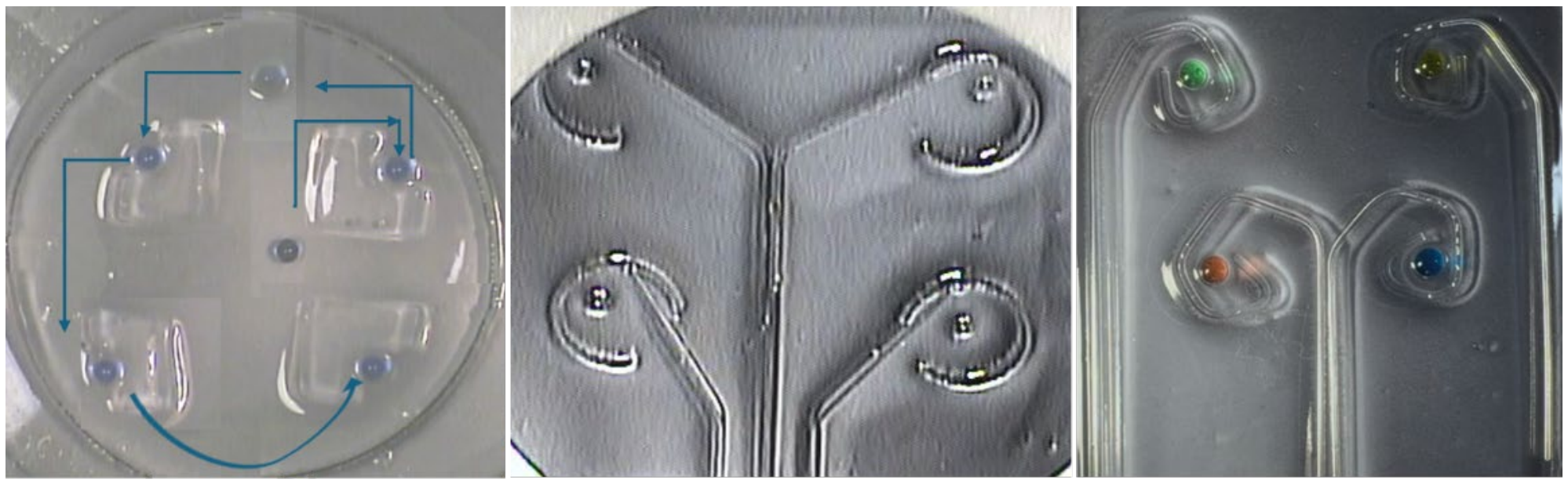

**Figure SM2**. Examples of surface-attached structures are shown here. In each experiment, 5 µl water droplets are steered using the electromagnetically controlled tilting platform (Figure 1). The PDMS substrate was first swollen in 5 cSt silicone oil. After draining the excess oil, the substrate was placed on the tilting device. The image displays four thin, L-shaped PDMS films attached to a plano-concave PDMS substrate using liquid PDMS as an adhesive. Arrows indicate the directions of water droplet motion. This figure was composed by joining pieces of multiple video frames.

**Steps on PDMS**

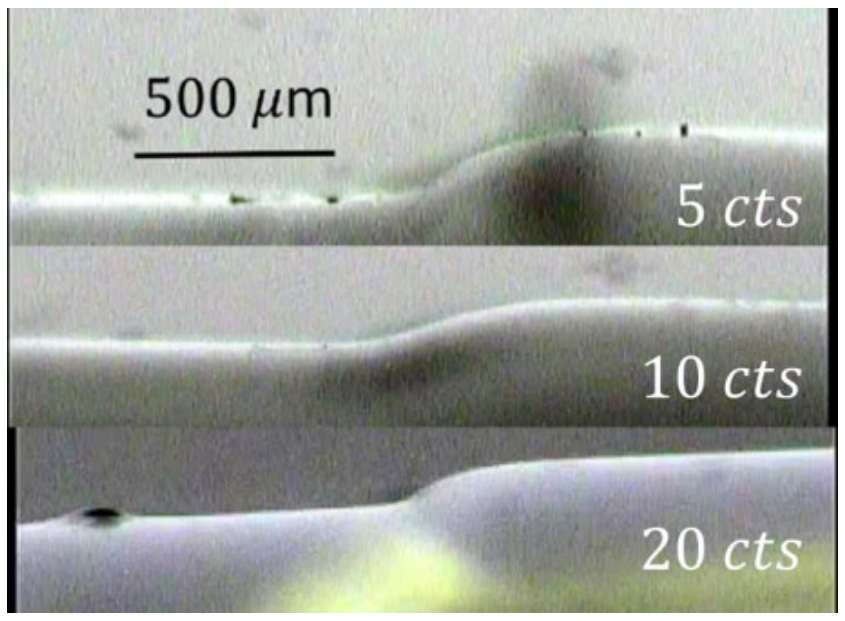

**Figure SM3.** Profiles of the edges formed on PDMS after the insertion of thin Si plates, followed by swelling in silicone fluids with viscosities of 5 cSt, 10 cSt, and 20 cSt, respectively.

Stepped structures similar to those outlined above can be generated by embedding a thin, rigid plate into the PDMS film. In this study, we used glass and silicon plates, which were securely bonded to the PDMS using the adhesion primer described in the experimental section. After sample preparation, the PDMS surface was coated with a thin layer of silicone oil. Upon swelling, a distinct microstructure emerged on the PDMS surface, visible in plan view. For glass strip inserts, optimal contrast was achieved with backlighting by placing the sample over a wire mesh. For silicon strip inserts, pattern development was visible along the edges of the strips.

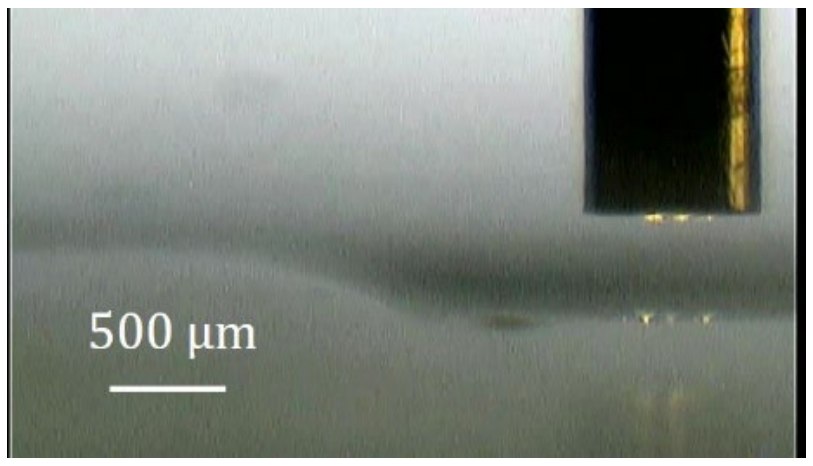

**Figure SM4.** Schematics of a Foton probe scanning step. The light emanating from the probe is reflected from the surface. The probe is initially positioned on the right side of the edge and moves left above the edge.

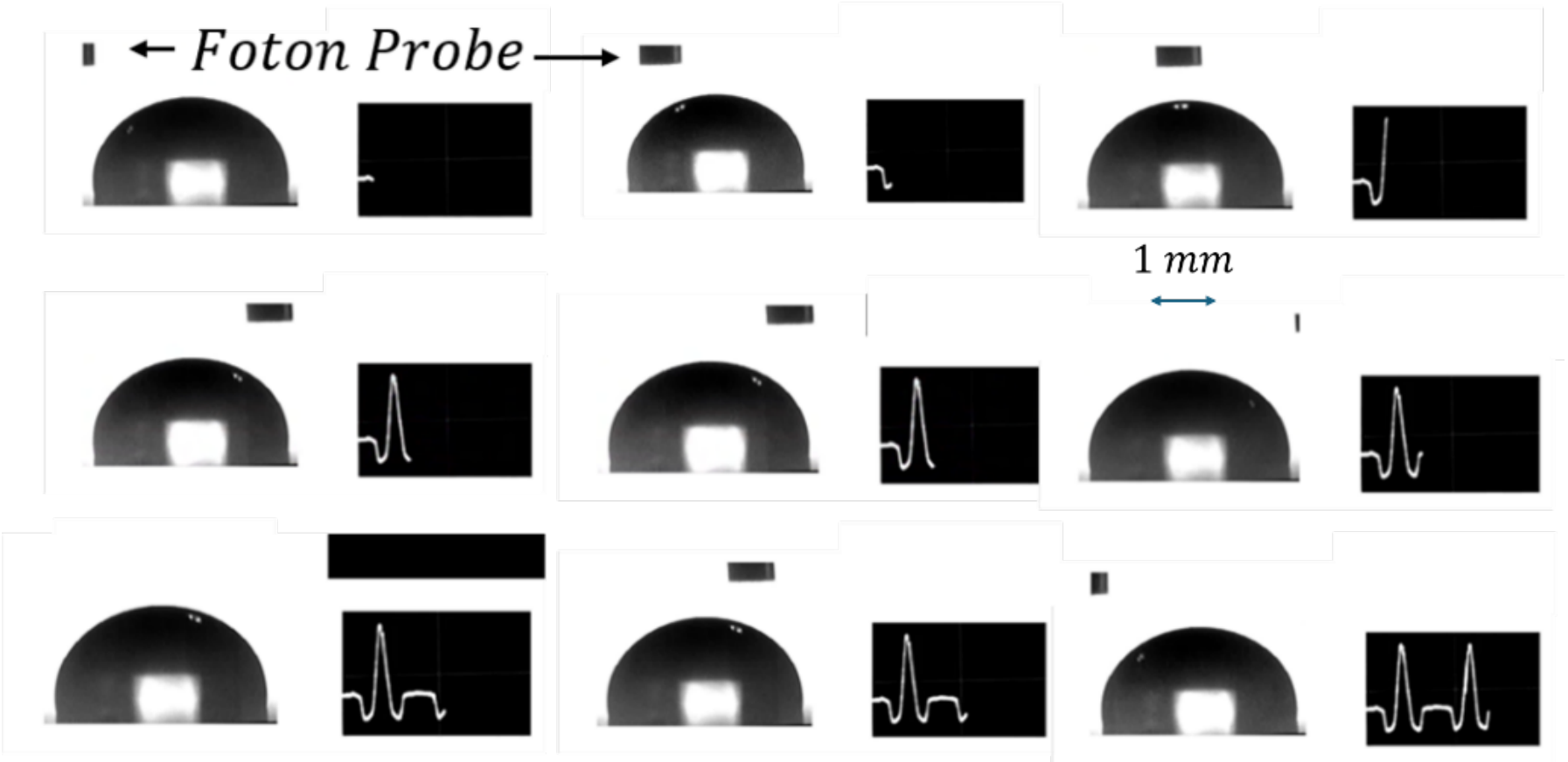

**Figure SM5**. This is an enlarged version of Figure 9 (Text). Here more video snapshots are added.

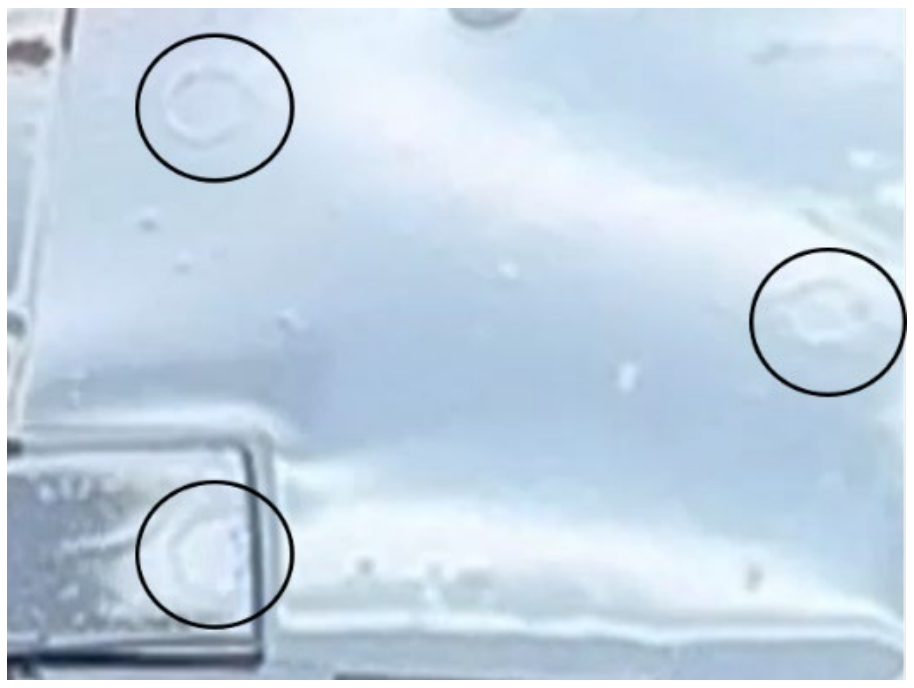

**Figure SM6**. Crater like deformation (diameter of the ring~1.5 mm) on a pre-swelled PDMS after placing a water drop on the surface and then removing it after 20 minutes of contact.

## Movie Captions

All the movies were processed using Adobe Express, Clipchamp, and an online video compressor (https://www.freeconvert.com/video-compressor) to improve clarity. The video sizes had to be reduced so they could be played in a reasonable amount of time. The original and current lengths are provided in the caption of each video.

**Movie S1**

Video of three bent capillaries attached to a PDMS substrate shows them serving as traps for a water droplet (see also Figure 1). A 5-μL colored water droplet was captured in the first trap on the left and then transported to the other traps by biaxially tilting the substrate on an electromagnetically controlled tilting platform. Original length 45s, current length 22s

**Movie S2**

Videos of water droplets moving through two surface-attached structures are shown here (see also Figure SM2). In each experiment, a 5 μl water droplet is steered using the electromagnetically controlled tilting platform (Figure 1). The video on the left shows a structure prepared by attaching four thin L-shaped PDMS films to a plano-concave PDMS substrate, using liquid PDMS as an adhesive. The other video shows morphed glass capillary tubes attached to flat PDMS films. Note how the water droplets cling to the straight and curved surface structures as they move through these complex features. The original length of the video on the left is 3m 50s; the current length is 28s. The original length of the video on the right is 4m 54s; the current length is 32s.

**Move S3**

Video demonstration of an acid-base reaction. Here, a 5-drop sample of 0.01N HNO3 is titrated by the sequential addition of 2-drop increments of aqueous NaOH (0.01N). As the acidity of the central drop decreases, its color changes from purple to green. Small drops are drawn into the large drops by capillary attraction, followed by fusion. See also Figure 5. Original video length 5m39 s, current length 15s.

**Movie S4**

Video of a 5 $\mu l$ blue-colored water droplet released near the edge of a plano-concave PDMS substrate. Note how it moves to the center of the substrate, sweeping silicone oil (5 cSt) on its way. When it reaches the center, the accumulated oil is clearly visible. When the substrate is When tilted, part of the accumulated oil moves with the droplet. The original length of the video 3m10s; current length 31s.

**Movie S5**

Directional motion of a water droplet (5 $\mu l$) on a plano-concave PDMS (swollen in 10 cSt silicone oil), illustrating uniaxial and biaxial motions via stepwise ($2^0$ at a time) counterclockwise rotation of a biaxial goniometer that controlled the tilt of the supporting stage. Original length of the video: 9m 56s, current length 52s.

**Movie S6**

Illustration of how the intensity of diffusely reflected light from a glycerol droplet varies with the probe's position above it. Here, a visually observed movie of the Foton probe is shown alongside the evolution of the Foton spectra as the probe moves from left to right and then back to its original position. The original video is 1m 58s long; the current length is 24s.

**Movie S7**

A blue water droplet ( $10\mu l$) moves along an inclined PDMS strip (swollen in 5 cSt silicone oil) until it reaches the edge. Initially, all droplets successfully cross the edge. However, after several droplets pass over the strip, a new droplet placed on the strip becomes pinned at the edge. When a droplet becomes pinned, it depins if a drop of silicone oil is placed behind it. The original video is 8m 26 s; the current version is 43s.n is 43s.

**Movie S8**

Demonstration of droplet manipulation on a plano-concave PDMS substrate using fluid diodes. By virtue of the orientation of the fluid diodes, a drop (e.g., the blue drop) placed at the lower arm can only move to the left, whereas a drop placed at the upper arm can only move to the right. A red drop in the middle has no restriction to move either way. Thus, when the substrate is tilted clockwise, the blue drop gets arrested at the fluidic gate midway in the lower arm, whereas the red drop continues to move. When the substrate is tilted counterclockwise, the blue drop gets arrested midway in the upper arm, whereas the one in the lower arm continues to move left along with the red drop between the two arms. Video's original length 14m 42s; current length 28s

**Movie S9**

Demonstration of manipulating multiple droplets on a flat PDMS substrate using fluid diodes, which allow small droplets to move in one direction. Here, the substrate was tilted in the range of $\pm 2^0$ to $\pm 5^0$. At these inclinations, the fused large droplets can overcome the barrier. Original length of this video 2m 36s; current length 39s.

**Movie S10**

Demonstration of a plano-concave PDMS platform to study a thermally activated reaction. As a proof of concept, we performed the well-known redox process used for glucose analysis, in which glucose reacts with Benedict's reagent at approximately 100 °C on a plano-concave PDMS substrate. One noteworthy feature is that the fused droplet tends to become pinned at the edge of the heating element due to thermal Marangoni pinning. However, as soon as this edge is overcome, the droplet runs away quickly. Original length of this video 10m 36s, current length 58s.